\documentclass
[twocolumn,showkeys,showpacs,nofootinbib]{revtex4}
\usepackage{amsfonts}
\usepackage{amsmath}
\usepackage{amsxtra}
\usepackage{graphicx}

\begin{document}


\title{Matter sources of gauge thermodynamics}


\author{Romualdo Tresguerres}
\email[]{romualdotresguerres@yahoo.es}
\affiliation{Instituto de F\'isica Fundamental\\
Consejo Superior de Investigaciones Cient\'ificas\\ Serrano 113
bis, 28006 Madrid, SPAIN}

\date{\today}

\begin{abstract}
A unified gauge approach to both, dynamics and thermodynamics involving gravity, is developed from the local realization of the Poincar\'e group as a particular instance of a spacetime group including translations. The formalism is applied to study the physical features of the spherical non-rotating matter source of an inner Schwarzschild metric.
\end{abstract}

\maketitle

\section{Introduction}

Dynamics and thermodynamics are inseparable aspects of macroscopic physical phenomena which demand a joint foundation. If, for completeness, the general relativistic dynamical view on spacetime is to be taken into account, then gravitation must also enter the scheme. A general self-consistent theory is required in which particular empirical laws should fit. Gauge theories of gravity deliver a suitable basis to construct such a fundamental approach merging the three theoretical elements mentioned above, as already discussed in \cite{Tresguerres:2013rnr}. Generally speaking, gauge theories, based on the local realization of symmetry groups, constitute a major formal tool for dealing with the physical interactions between world constituents. They provide a geometrized depiction of forces in the spirit of Klein's Erlangen Program of construction of geometries from suitable groups \cite{Sharpe:1996}. From a pure mathematical point of view, gauge theories consist in fiber bundles equipped with a structure group and with connections defining horizontality in a certain manifold \cite{Goeckeler}. Physically interpreted, connections are the carriers of interactions, whose dynamics follows from an action built on the bundle background.

In general, provided the field equations derived from the extremal condition on the action are fulfilled, the Noether theorems establish a necessary relationship between the symmetries of the action and some symmetry-associated conserved quantities \cite{Kosmann-Schwarzbach:2011}. In gauge theories, the Noether identities following from the symmetry invariance of the matter part of the action constitute the conservation laws of matter currents $J :={{\partial L^{\rm matt}}\over{\partial A}}$ defined as the derivatives of the matter Lagrangian with respect to the connections.

Among the physically relevant symmetry groups, spatial translations are well known to underly classical mechanics, being responsible for the conservation of momentum which is at the core of Newton's laws of motion \cite{Tresguerres:2007ih}. Thus, the natural way to construct a generalized approach to dynamics --either special or general relativistic-- seems to require invoking spacetime translations. However, no attention at all is paid to them when, as far as gravity is concerned, just general covariance is considered. The hidden contribution of spacetime translational symmetry only becomes apparent in the framework of certain gauge formulations of gravitation founded on the local realization of spacetime groups including translations, such as Poncar\'e Gauge Theories (PGT) and Metric-Affine Gravity (MAG) \cite{Hehl:1974cn}-\cite{Mielke:2017nwt}. The present paper is based on such gauge theories of gravity, where the usually ignored capital role played by spacetime translations is remarked. The key idea supporting our unifying view is that separate energy and momentum conservation equations are to be regarded as consequences of the invariance of the system's action (specifically of its matter part) under gauge translational transformations. In particular in PGT --the gauge theory having as structure group the Poincar\'e group, that is, the spacetime group of Special Relativity--, the place occupied by translations in the bundle structure is highlighted by the composite fiber bundle approach proposed by me in a previous paper \cite{Tresguerres:2002uh}, whose topological implications where studied by other authors in \cite{Cartwright:2016qfm}. In this framework, translations occur as a subgroup of the whole Poincar\'e gauge group in such a way that proper ({\it vertical}) translational gauge transformations are allowed along branched fibres (that is, fibres {\it bent} into mutually orthogonal translational and Lorentz fiber subspaces) attached to each single point of the base space. Composite bundles thus make it possible to treat spacetime translations as a gauge symmetry in the strict sense, clearly distinguished from base space diffeomorphisms. The canonical energy-momentum $\Sigma _\alpha$ plays the role of the corresponding standard conserved translational gauge current. Let us briefly comment this statement. (For a detailed discussion, see \cite{Tresguerres:2007ih}).

In analogy to the remaining forces as described by the Standard Model, in PGT gravitational interactions are carried by gauge connections, being these of two kinds, associated to the Lorentz and translational symmetries respectively. Lorentz connections do not differ significantly from ordinary gauge fields. However, regarding translational connections, their usual identification as tetrads is obscured by the fact that the latter ones transform as Lorentz covectors rather than as standard connections. Nonetheless, the apparent difficulty is clarified by considering tetrads as connections of a certain nonlinear kind. Actually, composite fiber bundles provide a natural geometric approach to nonlinear realizations of spacetime groups \cite{Tresguerres:2002uh}, according to which modified (nonlinear) translational connections possess the right transformation properties, allowing to regard them as tetrads, that is, as coframes $\{\vartheta ^\alpha\}$ of local cotangent spaces, playing a geometrical and a dynamical role at the same time. In exterior-calculus formulated gauge theories of arbitrary interactions in the presence of gravity, tetrads --as 1-form basis components-- are universally coupled to any physical quantity expressible as a differential form, while, on the other hand, being tetrads also translational connections, derivation of the sum of all matter Lagrangian pieces with respect to them defines energy-momentum $\Sigma _\alpha :={{\partial L^{\rm matt}}\over{\partial\vartheta ^\alpha}}$ as the conserved translative gauge current. The characterization of matter as source of gravity in PGT is completed by considering in addition the spin current associated to Lorentz symmetry, defined as the derivative of the matter Lagrangian with respect to the Lorentz connections, that is $\tau _{\alpha\beta} :={{\partial L^{\rm matt}}\over{\partial\Gamma ^{\alpha\beta}}}$. Actually, both, energy-momentum and spin current, play the role of matter sources in the gauge gravitational field equations, and the Noether identities provide conservation equations for them.

We assume these general results, derived in the framework of a Lagrangian approach, to hold also when dealing with phenomenological macroscopic matter lacking a well known Lagrangian. The form of the at first undetermined matter currents $\Sigma _\alpha$ and $\tau _{\alpha\beta}$ giving rise to a particular inner gravitational solution may then be inferred using the requirement of consistence with the field equations and the Noether identities as a guide. Once the sources are known, their dynamics (and thermodynamics) as established by the conservation equations can be studied separately.

The paper is organized as follows. In Sect.II, the fundamental results derived from PGT concerning the gravitational field equations and the Noether identities ensuring the conservation of energy-momentum and spin current are summarized. Sect.III is devoted to the foliation of the dynamical spacetime of gauge gravity, providing a suitable framework to define on it Eckart's projectors. The latter ones are used in Sect.IV to identify the physically meaningful covariant constitutive elements of the energy-momentum tensor, and in Sect.V to get, from the energy-momentum conservation equation established in PGT, separate conservation statements for both, scalar energy and Lorentz covariant momentum. In Sect.VI, we present a few examples of decomposition of the energy-momentum of Maxwell fields and of various kinds of fundamental matter into their constitutive elements in order to illustrate several features to be kept in mind in the following when considering macroscopic matter. Next I develop my main results, slightly differing from those already advanced in Ref.\cite{Tresguerres:2013rnr}. In Sect.VII, the energy conservation equation is supplemented with an interpretation of the quantities involved in it in terms of certain auxiliary variables in such a way that the former equation gives rise to the first law of thermodynamics. One of the auxiliary variables is the internal energy, regarded as a functional of the entropy and of other thermodynamically relevant quantities, so that its Lie derivative expands as the Gibbs fundamental equation of thermodynamics. The consistence condition between this alternative formulation of the first law and the previously derived one reveals itself as the second law of thermodynamics (with the only limitation that it does not necessarily predict entropy increase, which depends on the matter model considered). In Sect.VIII we introduce the Van der Waals fluid and the photon gas, described by variables satisfying both, suitable state equations and the Gibbs equation. Finally, in Sect.IX, the previous results are applied to analyze the thermodynamics and dynamics of a spherical, gaseous, radiating source of the gravitational field described by the inner Schwarzschild metric.

\section{Field equations and Noether identities in the gauge approach to gravity}

Similarly to ordinary gauge theories of local internal groups \cite{Sharpe:1996} \cite{Goeckeler}, the Lagrangian approach to Poincar\'e Gauge Theories (PGT) \cite{Hehl:1974cn}-\cite{Mielke:2017nwt} is obtained from a locally Poincar\'e invariant action
\begin{equation}
S = \int L\,\label{action}
\end{equation}
built from a Lagrangian density 4-form $L$ describing gravitation and matter. It consists of a functional of matter fields and their derivatives, along with the Poincar\'e connections necessary to ensure covariance --namely the tetrads $\vartheta ^\alpha$ (as nonlinear translational connections) and the Lorentz connections $\Gamma _\alpha{}^\beta$--, including in addition the corresponding field strengths \cite{Hehl:1995ue}, that is, the torsion
\begin{equation}
T^\alpha := D\,\vartheta ^\alpha = d\,\vartheta ^\alpha + \Gamma _\beta{}^\alpha\wedge\vartheta ^\beta\,,\label{torsiondef}
\end{equation}
and the Lorentz curvature
\begin{equation}
R_\alpha{}^\beta := d\,\Gamma _\alpha{}^\beta  + \Gamma _\gamma{}^\beta\wedge \Gamma _\alpha{}^\gamma\,,\label{curvdef}
\end{equation}
respectively. In fact, the Lagrangian density 4-form in (\ref{action}) can be split into a pure gravitational plus a matter piece
\begin{equation}
L= L^{\rm gr} + L^{\rm matt}\,,\label{totalLag}
\end{equation}
with respective functional dependence
\begin{eqnarray}
L^{\rm gr} &=& L^{\rm gr}(\,\vartheta ^\alpha\,,\,T^\alpha\,,\,R_\alpha{}^\beta\,)\,,\label{gravLag}\\
L^{\rm matt} &=& L^{\rm matt}(\,\vartheta ^\alpha\,,\Gamma ^{\alpha\beta};{\rm matter\hskip0.1cm variables}\,)\,.\label{mattLag}
\end{eqnarray}
The non-specified matter fields in (\ref{mattLag}) may include the gauge potentials of other interactions. Let us recall the general field equations and Noether identities \cite{Hehl:1974cn}-\cite{Mielke:2017nwt} obtained from (\ref{action}). (Here we do not pay attention to the field equations of the matter fields.) We define the gravitational energy-momentum 3-form
\begin{equation}
E_\alpha :={{\partial L^{\rm gr}}\over{\partial \vartheta ^\alpha}}\,,\label{defingrem}
\end{equation}
and the translative and Lorentzian excitation 2-forms
\begin{equation}
\quad H_\alpha :=-{{\partial L^{\rm gr}}\over{\partial T^\alpha}}\,,\quad H_{\alpha\beta}:=-\,{{\partial L^{\rm gr}}\over{\partial R^{\alpha\beta}}}\,,\label{definition02}
\end{equation}
as much as the translational and Lorentz matter currents already mentioned in the Introduction, that is, the energy-momentum 3-form
\begin{equation}
\Sigma _\alpha :={{\partial L^{\rm matt}}\over{\partial \vartheta ^\alpha}}\,,\label{definition03b}
\end{equation}
and the spin current 3-form
\begin{equation}
\tau _{\alpha\beta} :={{\partial L^{\rm matt}}\over{\partial \Gamma ^{\alpha\beta}}}\,,\label{definition03c}
\end{equation}
respectively. In terms of these quantities, varying (\ref{totalLag}) with respect to $\vartheta ^\alpha$ and $\Gamma ^{\alpha\beta}$ one gets the gravitational field equations
\begin{eqnarray}
DH_\alpha -E_\alpha&=&\Sigma _\alpha\,,\label{fieldeq1}\\
DH_{\alpha\beta} +\vartheta _{[\alpha }\wedge H_{\beta ]}&=&\tau _{\alpha\beta}\,,\label{fieldeq2}
\end{eqnarray}
being (\ref{fieldeq1}) a generalization of the Einstein equation and (\ref{fieldeq2}) a further fundamental equation absent from ordinary General Relativity, see \cite{Hehl:1995ue}. The sources of the gravitational fields are the energy-momentum $\Sigma _\alpha$ and the spin current $\tau _{\alpha\beta}$ respectively. Provided (\ref{fieldeq1}) and (\ref{fieldeq2}) --and the non specified matter field equations-- are fulfilled, the invariance of $L^{\rm matt}$ under Poincar\'e gauge transformations gives rise to the Noether conservation equations
\begin{eqnarray}
&&D\Sigma _\alpha =(\,e_\alpha\rfloor T^\beta )\wedge\Sigma _\beta +(\,e_\alpha\rfloor R^{\beta\gamma}\,)\wedge\tau _{\beta\gamma}\,,\label{sigmamattconserv}\\
&&D\tau _{\alpha\beta} +\vartheta _{[\alpha}\wedge\Sigma _{\beta ]}=0\,.\label{spincurrconserv}
\end{eqnarray}
The energy-momentum conservation law (\ref{sigmamattconserv}) is the Noether identity induced by the spacetime-translational invariance, while (intrinsic) angular momentum conservation (\ref{spincurrconserv}) follows from Lorentz invariance \cite{Tresguerres:2007ih} \cite{Hehl:1995ue}. The material energy-momentum in (\ref{sigmamattconserv}) is to be understood as the sum of all non-gravitational contributions to energy-momentum, including radiation as much as matter in the strict sense, so that no external forces due to interactions other than gravity manifest themselves. For instance, if all existing radiation were electromagnetic and described by Maxwell's theory, then we had to take $\Sigma _\alpha$ as the sum of matter and electromagnetic pieces
\begin{equation}
\Sigma _\alpha = \Sigma ^{\rm matt}_\alpha + \Sigma ^{\rm em}_\alpha\,,\label{mattemsum}
\end{equation}
for which, as shown by eqs. (19) and (22) of Ref.\cite{Tresguerres:2013rnr}, the following  Noether identities hold 
\begin{eqnarray}
D\,\Sigma ^{\rm matt}_\alpha &=& (\,e_\alpha\rfloor T^\beta )\wedge\Sigma ^{\rm matt}_\beta +(\,e_\alpha\rfloor R^{\beta\gamma}\,)\wedge\tau _{\beta\gamma}\nonumber\\
&&+(\,e_\alpha\rfloor F\,)\wedge J\,,\label{sigmamattconserv02}\\
D\,\Sigma ^{\rm em}_\alpha &=& (\,e_\alpha\rfloor T^\beta )\wedge\Sigma ^{\rm em}_\beta -(\,e_\alpha\rfloor F\,)\wedge dH\,,\label{sigmaemconserv02}
\end{eqnarray}
being $(\,e_\alpha\rfloor F\,)\wedge J =: f_\alpha$ in (\ref{sigmamattconserv02}) an external (electromagnetic) Lorentz force 4-form (with $F =dA$ as the electromagnetic field strength 2-form and $J$ as the electric current 3-form). Its structure is similar to that of the other (gravitational) {\it force} contributions in (\ref{sigmamattconserv02}), built with the field strengths and the matter currents of translational and Lorentz symmetry respectively. By adding up (\ref{sigmamattconserv02}) and (\ref{sigmaemconserv02}) taking into account the Maxwell equations
\begin{equation}
dH = J\,,\label{Maxwell01}
\end{equation}
expressed in terms of the electromagnetic excitation 2-form $H$, condition (\ref{sigmamattconserv}) is recovered, where the non-gravitational forces cancel out.

\section{Spacetime foliation and Eckart's projectors}

Relativistic continuous media are usually described with the help of the so called {\it hydrodynamic fourvelocity} and of Eckart's projectors defined in terms of it \cite{Eckart:1940te} \cite{DeGroot:1980dk}, allowing to deal separately with covariant timelike and spacelike quantities referred to a certain comoving frame. Here we introduce that approach bound to the foliation of the base space of Poincar\'e gauge gravity, being coframes defined on it at each point by tetrads. Let us consider a foliation of spacetime \cite{Hehl-and-Obukhov} induced by a 1-form $\omega = d\tau $ representing proper time, trivially satisfying Frobenius' foliation condition $\omega\wedge d\omega =0$. The timelike vector field $u$ such that
\begin{equation}
u\rfloor d\tau =1\,\label{utaucond01}
\end{equation}
represents a congruence of proper time oriented worldlines. Physical quantities expressed by $p$-forms $\alpha$ can be locally decomposed into longitudinal and transversal parts with respect to proper time as
\begin{equation}
\alpha = d\tau\wedge\alpha _{\bot} +\underline{\alpha}\,,\label{foliat1}
\end{equation}
being the longitudinal piece the projection of $\alpha$ along $u$
\begin{equation}
\alpha _{\bot} := u\rfloor\alpha\,,\label{long-part}
\end{equation}
and the transversal part a form whose components are projected on the orthogonal spatial directions as
\begin{equation}
\underline{\alpha}:= u\rfloor ( d\tau\wedge\alpha\,)\,.\label{trans-part}
\end{equation}
Evolution of a physical quantity in proper time is given by its Lie derivative along $u$, which, for $p$-forms, reads
\begin{equation}
{\it{l}}_u\alpha :=\,d\,(u\rfloor\alpha\,) + u\rfloor d\alpha\,.\label{Liederdef}
\end{equation}
The foliation of exterior derivatives of forms is performed in analogy to (\ref{foliat1}) as
\begin{equation}
d\,\alpha = d\tau\wedge\bigl(\,{\it{l}}_u\underline{\alpha} -\,\underline{d}\,\alpha _{\bot}\,\bigr) +\underline{d}\,\underline{\alpha}\,,\label{derivfoliat}\\
\end{equation}
expressed in terms of the Lie derivative (\ref{Liederdef}) and of $\underline{d}$ as the spatial differential. The foliation of local Lorentz coframes, consisting of tetrads, gives rise to Eckart's covariant projections on the proper time axis and the orthogonal spatial axes of the frame respectively \cite{Eckart:1940te} \cite{DeGroot:1980dk}. Being tetrads (when regarded as coframe components rather than as translative connections) a vector-valued one-form basis $\{\vartheta ^\alpha\}$ of the local cotangent space \cite{Hehl:1995ue}, the vector basis of the tangent space at each point --the {\it vierbein} frame $\{e_\alpha\}$-- is defined by the duality condition
\begin{equation}
e_\alpha\rfloor \vartheta ^\beta = \delta _\alpha ^\beta\label{dualitycond}
\end{equation}
with respect to the interior product. The splitting (\ref{foliat1}) of tetrads yields
\begin{equation}
\vartheta ^\alpha = d\tau\,u^\alpha + \underline{\vartheta}^\alpha\,,\label{tetradfoliat}
\end{equation}
where the longitudinal values
\begin{equation}
u^\alpha := u\rfloor\vartheta ^\alpha\label{fourvel}
\end{equation}
are the fourvelocity components of the time vector $u$ as evaluated in the local Lorentz frame $\{e_\alpha\}$; they are taken to satisfy
\begin{equation}
u^\alpha u_\alpha = -1\,\label{form01}
\end{equation}
for signature $o_{\alpha\beta} = {\rm diag}(-+++)$, in accordance with the timelike nature of $u$. From (\ref{dualitycond}) with (\ref{tetradfoliat}), that is, from
\begin{equation}
e_\alpha\rfloor \Big(\,d\tau\,u^\beta + \underline{\vartheta}^\beta\,\Bigr) = \delta _\alpha ^\beta\,,\label{dualitycondbis}
\end{equation}
one derives Eckart's projectors as follows. Let us rewrite condition (\ref{utaucond01}) --a particular instance of (\ref{Liederdef})-- referred to the local Lorentz frame $\{e_\alpha\}$ as
\begin{equation}
{\it l}_u\,\tau = u\rfloor d\tau = u^\alpha e_\alpha\rfloor d\tau =1\,.\label{utaucond02}
\end{equation}
Consistence between (\ref{form01}) and (\ref{utaucond02}) requires to assume that
\begin{equation}
e_\alpha \rfloor d\tau = -\,u_\alpha\,.\label{form02}
\end{equation}
Substituting (\ref{form02}) into (\ref{dualitycondbis}), one gets
\begin{equation}
e_\alpha\rfloor \underline{\vartheta}^\beta = h_\alpha{}^\beta\,,\label{dualitycondbisbis}
\end{equation}
expressed in terms of the symmetric tensor defined as
\begin{equation}
h_\alpha{}^\beta :=\delta _\alpha ^\beta + u_\alpha u^\beta\,.\label{projectr01}
\end{equation}
This tensor is a projector with the properties
\begin{eqnarray}
h_\alpha{}^\mu h_\mu{}^\beta &=& h_\alpha{}^\beta\,,\label{projectr02}\\
h_\alpha{}^\beta u_\beta &=& 0\,,\label{projectr02}\\
h_\mu{}^\mu &=& 3\,.\label{projectr03}
\end{eqnarray}
Actually, (\ref{projectr01}) is the {\it spatial projection tensor} which selects the tensorial contributions which are transversal to proper time. For instance, when contracted with a fourvector, the latter's part parallel to the fourvelocity $u^\alpha$ is obliterated in view of (\ref{projectr02}). Consequently, using (\ref{projectr01}) and its complementary longitudinal projector $-u_\alpha u^\beta$, any fourvector $F^\alpha$ can be locally decomposed into timelike and spacelike elements as
\begin{equation}
F^\alpha \equiv \bigl( -u^\alpha u_\beta  +h^\alpha{}_\beta\bigr)F^\beta =: u^\alpha f +f^\alpha\,,\label{Fdecomp01}
\end{equation}
where Lorentz covariance is preserved separately for both pieces. The projection of $F^\alpha$ along the time direction is a scalar quantity $f:=-u_\beta F^\beta $, while the covariant contribution $f^\alpha := h^\alpha{}_\beta F^\beta$ representing the spatial vector part is orthogonal to the proper time components $u^\alpha$ since $f^\alpha u_\alpha =0$ (and thus possesses only three independent degrees of freedom). Notice that, comparing (\ref{tetradfoliat}) with $\vartheta ^\alpha \equiv \bigl( -u^\alpha u_\beta  +h^\alpha{}_\beta\bigr)\vartheta ^\beta$, it follows
\begin{eqnarray}
d\tau &=& - u_\beta\,\vartheta ^\beta\,,\label{dtau}\\
\underline{\vartheta}^\alpha &=& h^\alpha{}_\beta\,\vartheta ^\beta\,.\label{transtetrad}
\end{eqnarray}
In hydrodynamics, it is usual to fix the local Lorentz frame to be at rest either with respect to the particle diffusion flow (Eckart) or to the energy flow (Landau-Lifshitz), or in any other way, so that certain frame related quantities vanish \cite{Eckart:1940te} \cite{DeGroot:1980dk} \cite{Minami:2012hs}. We prefer to let the rest frame unspecified in order not to restrict the mathematical generality of the treatment. Complementarily to the foliation (\ref{tetradfoliat}) of the tetrads, the definition of the eta basis and its foliations, to be used extensively in the following, can be found in Appendix A.

\section{Basic constituents of the energy-momentum}

In gauge theories, connections (gauge fields) describing interactions are required in order to preserve the invariance of the action and the covariance of the field equations under local symmetry transformations. A singular feature of tetrads which distinguishes them from other connections is that they couple non-minimally (let us say, universally) to the remaining gauge fields as much as to all physical quantities represented by differential forms. A main consequence of this fact is that energy-momentum (\ref{definition03b}), as the translational current obtained by deriving the matter Lagrangian with respect to tetrads, is an unifying theoretical tool containing information about all matter fields and interactions, thus deserving special attention. In the present Section, we are going to identify the constitutive elements of the energy-momentum by decomposing it covariantly with the help of Eckart's projectors.

Projection of components along the frame axes as shown in (\ref{Fdecomp01}) for fourvectors generalizes straightforwardly to tensorial quantities. Let us consider the energy-momentum 3-form built with the eta basis element (\ref{antisym3form})
\begin{equation}
\Sigma _\alpha = \Sigma _\alpha{}^\beta\,\eta _\beta\,,\label{enmom01}
\end{equation}
whose tensorial components $\Sigma _\alpha{}^\beta$ are in general non-symmetric, in accordance with Eqn.(\ref{spincurrconserv}) for spin current conservation. Making use of Eckart's projectors \cite{Eckart:1940te} \cite{Muschik:2015oua}, we define the proper constituents
\begin{eqnarray}
\rho &:=& u^\mu\Sigma _\mu{}^\nu u_\nu\,,\label{rho01}\\
q^\beta &:=& u^\mu\Sigma _\mu{}^\nu  h_\nu{}^\beta\,,\label{flux01}\\
p_\alpha &:=& h_\alpha{}^\mu\Sigma _\mu{}^\nu u_\nu\,,\label{momdensity01}\\
p h_\alpha{}^\beta + \Pi _\alpha{}^\beta &:=& h_\alpha{}^\mu\Sigma _\mu{}^\nu h_\nu{}^\beta\,.\label{pressandS01}
\end{eqnarray}
In terms of them, the general energy-momentum tensor, able to describe any kind of matter, reads
\begin{equation}
\Sigma _\alpha{}^\beta = \rho\,u_\alpha u^\beta + p\,h_\alpha{}^\beta -u_\alpha q^\beta -p_\alpha u^\beta +\Pi _\alpha{}^\beta\,.\label{enmom02}
\end{equation}
The energy density (\ref{rho01}) is a scalar quantity. Energy flux (\ref{flux01}) and momentum density (\ref{momdensity01}) are spacelike vectors such that $u_\beta q^\beta =0$ and $u^\alpha p_\alpha =0$, so that, in spite of their Lorentz covariant form, each one posseses only three independent degrees of freedom. For them, the following relations also hold
\begin{eqnarray}
q^\beta &=& \rho u^\beta + u^\mu\Sigma _\mu{}^\beta\,,\label{flux03bis}\\
p_\alpha &=& \rho u_\alpha +\Sigma _\alpha{}^\nu u_\nu\,.\label{momdens02}
\end{eqnarray}
Their values coincide if the energy-momentum tensor (\ref{enmom02}) is symmetric, but their physical roles differ in any case. (That is best shown in the exterior calculus formulation, see (\ref{energy01}) and (\ref{enmom03}) below.) The total stress tensor (\ref{pressandS01}) includes a separate pressure contribution, being $p$ a scalar quantity. The proper stress tensor $\Pi _\alpha{}^\beta$, such that $u^\alpha\Pi _\alpha{}^\beta =0$ and $\Pi _\alpha{}^\beta u_\beta =0$, is not symmetric in general, but decomposes as
\begin{equation}
\Pi _{\alpha\beta} = \Pi _{(\alpha\beta )} + \Pi _{[\alpha\beta ]}\,.\label{Stresstens01}
\end{equation}
Its symmetric part
\begin{equation}
S_{\alpha\beta} := \Pi _{(\alpha\beta )}\,,\label{Stresstens02}
\end{equation}
called the viscous stress tensor, divides into shear and bulk viscosities. The latter one consists of the trace $S_\mu{}^\mu$, while shear viscosity is defined as the traceless part of (\ref{Stresstens02}), that is
\begin{equation}
{\slash\!\!\!\!S}_\alpha{}^\beta := S_\alpha{}^\beta -{1\over 3}\,h_\alpha{}^\beta S_\mu{}^\mu\,.\label{tracefreeS01}
\end{equation}
The trace of (\ref{enmom02}) is found to be
\begin{equation}
\Sigma _\mu{}^\mu = -\rho +3p +S_\mu{}^\mu\,.\label{Strace01}
\end{equation}
Its value depends essentially on the mass content of the material fields it is built of, as will be seen later.

\section{Separate energy and momentum conservation}

Eckart's spatial projectors (\ref{projectr01}), together with the complementary temporal ones, namely $-u_\alpha u ^\beta$, allow to split the energy-momentum conservation equation (\ref{sigmamattconserv}) into separate conservation equations for energy and momentum, as the Noether identities corresponding to time and space gauge invariance respectively. In order to perform the splitting, we first decompose the energy-momentum 3-form (\ref{enmom01}) as
\begin{eqnarray}
\Sigma _\alpha &&\equiv ( -u_\alpha u^\beta + h_\alpha{}^\beta ) \Sigma _\beta\nonumber\\
&&=: u_\alpha\,\epsilon +\widetilde{\Sigma}_\alpha\,,\label{enmom02bis}
\end{eqnarray}
being the energy and momentum current 3-forms respectively defined as
\begin{eqnarray}
\epsilon &:=& -u^\beta\,\Sigma _\beta\,,\label{energycurr01}\\
\widetilde{\Sigma}_\alpha &:=& h_\alpha{}^\beta\,\Sigma _\beta\,.\label{momcurr01}
\end{eqnarray}
Notice the scalar character \cite{Eckart:1940te} of the energy (\ref{projectr01}). By foliating the latter and the momentum (\ref{momcurr01}) according to (\ref{foliat1}), one gets
\begin{eqnarray}
\epsilon &=& d\tau\wedge\epsilon _{\bot} +\underline{\epsilon}\,,\label{eneryfoliat1}\\
\widetilde{\Sigma}_\alpha &=& d\tau\wedge\widetilde{\Sigma}_{\alpha\bot} + \underline{\widetilde{\Sigma}}_\alpha\,,\label{mom01foliat}
\end{eqnarray}
or explicitly, as calculated from (\ref{enmom01}) with (\ref{enmom02}) and (\ref{eta03}),
\begin{eqnarray}
\epsilon &=& d\tau\wedge q^\beta\overline{\eta}_\beta +\rho\overline{\eta}\,,\label{energy01}\\
\widetilde{\Sigma}_\alpha &=& -d\tau\wedge\bigl(\,p\,\overline{\eta}_\alpha + \Pi _\alpha{}^\beta\,\overline{\eta}_\beta\bigr) - p_\alpha\overline{\eta}
\,,\label{enmom03}
\end{eqnarray}
showing the material energy and momentum fluxes and densities in terms of (\ref{rho01})-(\ref{pressandS01}) to be
\begin{eqnarray}
\epsilon _{\bot} &=& q^\beta\overline{\eta}_\beta\,,\label{eneryflux01}\\
\underline{\epsilon} &=& \rho\overline{\eta}\,,\label{eneryamount01}\\
\widetilde{\Sigma}_{\alpha\bot} &=& -\bigl(\,p\,\overline{\eta}_\alpha + \Pi _\alpha{}^\beta\,\overline{\eta}_\beta\bigr)\,,\label{momflux01}\\
\underline{\widetilde{\Sigma}}_\alpha &=& - p_\alpha\overline{\eta}\,.\label{momamount01}
\end{eqnarray}
The spatial 2-form (\ref{eneryflux01}) measures the flow of energy through a surface element per unit time, and the 3-form (\ref{eneryamount01}) represents the amount of energy stored in a volume element. Analogously, (\ref{momflux01}) is the momentum flux built from pressure and stress tensor contributions, and (\ref{momamount01}) is the momentum associated to an elementary volume.

The distinguished conservation equations for energy and momentum are calculated as follows. Contracting (\ref{sigmamattconserv}) with $u^\alpha$ and taking into account the definition (\ref{energycurr01}) of the energy current 3-form, we get the gauge equation for energy conservation
\begin{equation}
d\,\epsilon + {\cal \L\/}_u\,\vartheta ^\alpha\wedge\Sigma _\alpha + R_{\bot}^{\alpha\beta}\wedge\tau _{\alpha\beta} =0\,,\label{mattender01}
\end{equation}
involving the Lie derivative (\ref{thetaLiederiv01}) of the tetrads. On the other hand, multiplying (\ref{sigmamattconserv}) by the spatial projector (\ref{projectr01}) and using the decomposition (\ref{enmom02bis}), it follows
\begin{eqnarray}
&&h_\alpha{}^\beta \Bigl[ D\widetilde{\Sigma}_\beta -(\,e_\beta\rfloor T^\mu )\wedge\widetilde{\Sigma}_\mu -(\,e_\beta\rfloor R^{\mu\nu}\,)\wedge\tau _{\mu\nu} \Bigr]\nonumber\\
&&\hskip0.3cm + \Bigl[ D u_\alpha - h_\alpha{}^\beta \bigl( e_\beta\rfloor T^\mu\bigr) u_\mu \Bigr]\wedge\epsilon =0\,.\label{mattmomder01}
\end{eqnarray}
Next we foliate (\ref{mattender01}) making use of (\ref{energy01}), (\ref{enmom03}), (\ref{thetaLiederiv02}) and (\ref{volLiederiv02}), and taking into account that, according to (\ref{derivfoliat}), the derivative of the energy (\ref{energy01}) reads
\begin{equation}
d\epsilon = d\tau\wedge \bigl[\,{\it l}_u\bigl(\rho\overline{\eta}\bigr) -\underline{d}\,\bigl( q^\beta\overline{\eta}_\beta \bigr) \bigr]\,.\label{energyderivfol01}
\end{equation}
From (\ref{mattender01}) then follows
\begin{eqnarray}
{\it l}_u (\rho\overline{\eta}) &-& \underline{d}\,(q^\beta\overline{\eta}_\beta ) - p_\alpha\,{\cal \L\/}_u u^\alpha\,\overline{\eta} - T_{\bot}^\alpha\,u_\alpha \wedge q^\beta\overline{\eta}_\beta\nonumber\\
&+& p\,{\it l}_u \overline{\eta} +{\cal \L\/}_u \underline{\vartheta}^\alpha\wedge\Pi _\alpha{}^\beta\overline{\eta}_\beta - R_{\bot}^{\alpha\beta}\wedge{(\tau _{\alpha\beta})}_{\bot} =0\,,\nonumber\\
\label{mattender03}
\end{eqnarray}
constituting the most general form of the energy conservation law in terms of the constitutive elements of (\ref{enmom02}). Analogously, using (\ref{covderfol01}) to foliate the covariant derivative of the momentum as
\begin{equation}
D\widetilde{\Sigma}_\alpha = d\tau\wedge\bigl( {\cal \L\/}_u \underline{\widetilde{\Sigma}}_\alpha -\underline{D}\widetilde{\Sigma}_{\alpha\bot}\bigr)\,,\label{momcovderivfol01}
\end{equation}
from (\ref{mattmomder01}) with (\ref{energy01}), (\ref{enmom03}), (\ref{surfder02}) and (\ref{accel01}) we find
\begin{eqnarray}
&&h_\alpha{}^\beta \Bigl[ -{\cal \L\/}_u\bigl( p_\beta\,\overline{\eta}\bigr) + \underline{D}\bigl(\Pi _\beta{}^\gamma\,\overline{\eta}_\gamma\bigr)\Bigr] +\underline{d}p\wedge\overline{\eta}_\alpha\nonumber\\
&&\hskip0.4cm -\bigl( e_\alpha\rfloor T_{\bot}^\mu \bigr)\, p_\mu\,\overline{\eta} -\bigl( e_\alpha\rfloor\underline{T}^\mu \bigr)\wedge\Pi _\mu{}^\gamma \,\overline{\eta}_\gamma\nonumber\\
&&\hskip0.4cm +( e_\alpha\rfloor R_{\bot}^{\mu\nu}\,)\underline{\tau}_{\mu\nu}+( e_\alpha\rfloor\underline{R}^{\mu\nu}\,)\wedge(\tau _{\mu\nu})_{\bot}\nonumber\\
&&\hskip0.3cm = \Bigl[\underline{D} u_\alpha -\bigl( e_\alpha\rfloor\underline{T}^\mu\bigr) u_\mu\Bigr]\wedge q^\beta\,\overline{\eta}_\beta\,.\label{mattmomder04}
\end{eqnarray}
Eq.(\ref{mattmomder04}) is a general relativistic version of the Navier-Stokes dynamical law. In the following, we will use the conservation laws (\ref{mattender03}) and (\ref{mattmomder04}) derived in the framework of gauge gravity as a guide to develop a gauge covariant approach to fluid dynamics and thermodynamics.

\section{Energy-momentum of different kinds of Maxwell fields and fundamental matter}

In PGT, local translational invariance underlies both, the gravitational and the material Lagrangian pieces (\ref{totalLag}) used to build the action (\ref{emLagrang1}), even if that symmetry, in the case of $L^{\rm matt}$, may be not apparent. Indeed, all kinds of matter contribute to energy-momentum due to their coupling to translations. Let us illustrate the diverse forms of energy-momentum with a few examples concerning electromagnetic fields and fundamental matter.

\subsection{Free electromagnetic radiation}

For Maxwell fields in free space (without polarized nor magnetized matter), we first consider \cite{Hehl-and-Obukhov} \cite{Obukhov:2003cc} \cite{Obukhov:2008dz} a piece of the matter Lagrangian density 4-form (\ref{totalLag}) built from electromagnetic fields as
\begin{equation}
L^{\rm em}=-{1\over 2}\,F\wedge\,^*F\,,\label{emLagrang1}
\end{equation}
with $F = dA$ as the electromagnetic field strength 2-form. Its contribution to energy-momentum (\ref{definition03b}) is due to the hidden non-minimal coupling of any differential form, in particular ${}^*F$, to tetrads (that is, recall once more, to nonlinear translational connections) as ${}^*F ={1\over 2}\,\eta ^{\alpha\beta}( e_\beta\rfloor e_\alpha\rfloor F )$, see (\ref{dualform}). The role played by tetrads is revealed by the variational formula (\ref{dualvar}), allowing to calculate the canonical electromagnetic energy-momentum
\begin{equation}
\Sigma ^{\rm em}_\alpha = {1\over 2}\,\left[\,\left( e_\alpha\rfloor F\right)\wedge {}^*F -F\wedge\left( e_\alpha\rfloor {}^*F\right)\,\right]\,.\label{emenergymom01}
\end{equation}
Foliating the electromagnetic strength and its Hodge dual, we get respectively
\begin{eqnarray}
F &=& -d\tau\wedge E + B = -d\tau\wedge E_\mu\underline{\vartheta}^\mu + B^\mu\overline{\eta}_\mu\,,\label{emfoliat01}\\
{}^*F &=& d\tau\wedge {}^{\#}B + {}^{\#}E = d\tau\wedge B_\mu\underline{\vartheta}^\mu + E^\mu\overline{\eta}_\mu\,,\label{emfoliat02}
\end{eqnarray}
where we introduced the electric and magnetic field fourvector components $E_\mu$ and $B_\mu$ as convenient variables, with their number of independent degrees of freedom restricted by conditions $u_\mu E^\mu =0$ and $u_\mu B^\mu =0$. Replacing (\ref{emfoliat01}) and (\ref{emfoliat02}) in (\ref{emenergymom01}), a straightforward calculation yields an explicit form for (\ref{energy01}) and (\ref{enmom03}), allowing to identify the constitutive elements of the tensorial components (\ref{enmom02}). The latter ones result to be symmetric (that is, with $p_\alpha = q_\alpha$ and $\Pi _{\alpha\beta} = S_{\alpha\beta}$), being
\begin{eqnarray}
\rho &=& {1\over 2}\Bigl( E_\mu E^\mu + B_\mu B^\mu\Bigr)\,,\label{emrho01}\\
q_\alpha &=& -\epsilon _{\alpha\mu\nu}\,E^\mu B^\nu\,,\label{emflux01}\\
p &=& {1\over 6}\Bigl( E_\mu E^\mu + B_\mu B^\mu\Bigr)\,,\label{empress01}\\
S_\alpha{}^\beta &=& -\Bigl( E_\alpha E^\beta + B_\alpha B^\beta\Bigr) + {1\over 3}\,h_\alpha{}^\beta \Bigl( E_\mu E^\mu + B_\mu B^\mu\Bigr)\,.\nonumber\\
\label{emsymmS01}
\end{eqnarray}
The Poynting vector (\ref{emflux01}) plays a double role as both, energy flux and momentum density. From (\ref{emrho01}) and (\ref{empress01}) follows the relation
\begin{equation}
p = {1\over 3}\,\rho\,\label{emeffpress01}
\end{equation}
between pressure and energy density, and from (\ref{emsymmS01}) we find the bulk viscosity to vanish
\begin{equation}
S_\mu{}^\mu =0 \,,\label{zerobulk01}
\end{equation}
so that, replacing (\ref{emeffpress01}) and (\ref{zerobulk01}) in (\ref{Strace01}) we get
\begin{equation}
\Sigma _\mu{}^\mu = 0\,.\label{ememtrace01}
\end{equation}
The zero trace (\ref{ememtrace01}) of the energy-momentum tensor, as much as the previous results (\ref{emeffpress01}) and (\ref{zerobulk01}), may be generalized as common features of all non-massive fields.

\subsection{Electromagnetic radiation in material media}

The macroscopic Maxwell equations (\ref{Maxwell01}) in material media admit two alternative treatments \cite{Tresguerres:2013rnr} \cite{Hehl-and-Obukhov} \cite{Obukhov:2008dz} in terms of the bare electromagnetic excitation 2-form $H^{\rm bare} =\,{}^*F$ and the matter excitation 2-form $H^{\rm matt} = - d\tau\wedge M +P$ built from the magnetization 1-form $M$ and the polarization 2-form $P$ characteristic for each medium. The two approaches differ in wether one takes $H^{\rm matt}$ as part of the fields, or uses it instead to modify the free electric current $J^{\rm free} = -d\tau\wedge j + \rho _{\rm el}\overline{\eta}$ by adding to it bound charges and currents. That is, either one considers
\begin{equation}
d\bigl( H^{\rm bare} + H^{\rm matt}\bigr)= J^{\rm free}\,,\label{Maxwell02}
\end{equation}
with
\begin{eqnarray}
H^{\rm tot} := H^{\rm bare} + H^{\rm matt} &=& d\tau\wedge\bigl({}^\#B -M\bigr) + \bigl({}^\#E +P\bigr)\nonumber\\
&=:& d\tau\wedge {\cal H} + {\cal D}\,,\label{Htot01}
\end{eqnarray}
or one uses $dH^{\rm matt}$ in (\ref{Maxwell02}) to redefine the electric current 3-form as
\begin{equation}
dH^{\rm bare} = J^{\rm free} - dH^{\rm matt}\,,\label{Maxwell03}
\end{equation}
with
\begin{equation}
J^{\rm tot} := J^{\rm free} - dH^{\rm matt} = -d\tau\wedge ( j +{\it{l}}_u P +\underline{d}\,M\,) + (\rho _{\rm el}\overline{\eta} -\underline{d}\,P\,)\,.\label{totcurr}
\end{equation}
The total current (\ref{totcurr}) is the sum of the free one $J^{\rm free}$ and bound contributions due to the material polarizable and magnetizable medium. 

In the second case, the energy momentum is built from $H^{\rm bare} =\,{}^*F$, according to the Maxwell-Lorentz electromagnetic spacetime relation for the excitation, thus being identical with the symmetric quantity (\ref{emenergymom01}) already studied. Let us now pay attention to the first case. The canonical (Minkowski) \cite{Obukhov:2008dz} energy-momentum 3-form constructed from (\ref{Htot01}) reads
\begin{equation}
\Sigma ^{\rm em}_\alpha = {1\over 2}\,\left[\,\left( e_\alpha\rfloor F\right)\wedge H^{\rm tot} -F\wedge\left( e_\alpha\rfloor H^{\rm tot}\right)\,\right]\,.\label{emenergymom02}
\end{equation}
As we did in (\ref{emfoliat01}) and (\ref{emfoliat02}) for the electric and magnetic fields, we write the longitudinal and transversal pieces of the electromagnetic excitation (\ref{Htot01}) in terms of fourvector components as
\begin{equation}
H^{\rm tot} = d\tau\wedge{\cal{H}} +{\cal{D}} = d\tau\wedge {\cal{H}}_\mu\underline{\vartheta}^\mu + {\cal{D}}^\mu\overline{\eta}_\mu\,.\label{Hquant01}
\end{equation}
The number of independent degrees of freedom of both, the magnetic field intensity vector ${\cal{H}}_\mu$ and the electric displacement vector ${\cal{D}}^\mu$, are reduced by conditions $u^\mu{\cal{H}}_\mu =0$  and $u_\mu{\cal{D}}^\mu =0$ respectively. The physical quantities in the rhs of (\ref{enmom02}) as calculated from (\ref{emenergymom02}) and (\ref{Hquant01}) are found to be
\begin{eqnarray}
\rho &=& {1\over 2}\Bigl( E_\mu {\cal{D}}^\mu + B_\mu {\cal{H}}^\mu\Bigr)\,,\label{emrho01bis}\\
q^\beta &=& -\epsilon ^\beta{}_{\mu\nu}\,E^\mu {\cal{H}}^\nu\,,\label{emflux01bis}\\
p_\alpha &=& -\epsilon _{\alpha\mu\nu}\,{\cal{D}}^\mu B^\nu\,,\label{emmomdens01bis}\\
p &=& {1\over 6}\Bigl( E_\mu {\cal{D}}^\mu + B_\mu {\cal{H}}^\mu\Bigr)\,,\label{empress01bis}\\
\Pi _\alpha{}^\beta &=& -\Bigl( E_\alpha {\cal{D}}^\beta + B_\alpha {\cal{H}}^\beta\Bigr) + {1\over 3}\,h_\alpha{}^\beta \Bigl( E_\mu {\cal{D}}^\mu + B_\mu {\cal{H}}^\mu\Bigr)\,.\nonumber\\
\label{emsymmS01bis}
\end{eqnarray}
The results (\ref{emeffpress01})-(\ref{ememtrace01}) remain valid, but notice that (\ref{enmom02}) is not symmetric. In particular, energy flux (\ref{emflux01bis}) and momentum density (\ref{emmomdens01bis}) are different. This remains true even in the simplest linear case when
\begin{equation}
{\cal{D}}=\varepsilon\,{}^{\#}E\,,\qquad {\cal{H}}= {1\over{\mu}}\,{}^{\#}B\,,\label{Hquant02}
\end{equation}
being $\varepsilon\mu ={1\over{v^2}}$, with $v$ as the phase velocity of light in the medium. Comparing (\ref{emflux01bis}) and (\ref{emmomdens01bis}) we get
\begin{equation}
p_\alpha = -\epsilon _{\alpha\mu\nu} {\cal{D}}^\mu B^\nu = -\varepsilon\mu\,\epsilon _{\alpha\mu\nu} E^\mu {\cal{H}}^\nu = \varepsilon\mu\,q_\alpha\,,\label{fluxmomrel}
\end{equation}
so that the equality between $p_\alpha$ and $q_\alpha$ only holds when $\varepsilon\mu ={1\over{c^2}}$ with $c=1$, that is, when the phase velocity of light is that of light in vacuum rather than in a medium. The difficulties concerning the asymmetry of the electromagnetic energy-momentuma tensor in material media, and the controversy on the choice of the most convenient among various energy-momenta to describe phenomena, are discussed for instance in \cite{Obukhov:2003cc} and \cite{Obukhov:2008dz}.

\subsection{Fundamental fermionic matter}

A less problematic example of an essentially non-symmetric energy-momentum tensor is found in the context of fermions as described by the Dirac Lagrangian \cite{Mielke:1997cb} written as
\begin{equation}
L^{\rm Dir}={i\over 2}\,(\,\overline{\psi}\,\,^*\gamma\wedge D\psi +\overline{D\psi}\wedge\,^*\gamma\psi\,) -\,^*m\,\overline{\psi}\psi\,,\label{DiracLagrang1}
\end{equation}
built from Poincar\'e $\otimes$ $U(1)$ (or Lorentz $\otimes$ $U(1)$, see \cite{Tiemblo:2005sx}) covariant derivatives, where $\gamma :=\vartheta ^\alpha\,\gamma _\alpha\,$, with $\gamma ^\alpha$ as the Dirac gamma matrices, so that $\,{}^*\gamma :=\eta ^\alpha\,\gamma _\alpha \,$; see (\ref{antisym3form}). In this case, a non-trivial spin current exists
\begin{equation}
\tau _{\alpha\beta}=-{1\over 2}\,\overline{\psi}\,\left(\,
\sigma _{\alpha\beta}\,{}^*\gamma +\,{}^*\gamma\,{}\sigma _{\alpha\beta}\,\right)\,\psi\,,\label{explicitspincurr}
\end{equation}
whose conservation equation (\ref{spincurrconserv}) involves antisymmetric energy-momentum contributions. The energy-momentum 3-form (\ref{definition03b}) derived from (\ref{DiracLagrang1}) reads
\begin{equation}
\Sigma ^{\rm Dir}_\alpha = {i\over 2}\Bigl[(\,e_\alpha\rfloor \,\overline{\psi}\,\,{}^*\gamma\,)\wedge D\psi -\overline{D\psi}\wedge (\,e_\alpha\rfloor \,{}^*\gamma\,\psi\,)\Bigr] - m\overline{\psi}\psi\,\eta _\alpha\,,\label{Direnergymom01}
\end{equation}
for which one calculates its constitutive elements (\ref{enmom02}) to be
\begin{eqnarray}
\rho &=& m\,\overline{\psi}\psi + S_\mu{}^\mu\,,\label{Dirrho01}\\
q^\beta &=& {i\over 2}\Bigl(\overline{\psi}\gamma ^\mu \,{\cal \L\/}_u\psi -\overline{{\cal \L\/}_u\psi}\,\gamma ^\mu\psi\Bigr) h_\mu{}^\beta\,,\label{Dirflux01}\\
p &=& {i\over 2}\Bigl(\overline{\psi}\gamma ^\mu \,{\cal \L\/}_u\psi -\overline{{\cal \L\/}_u\psi}\,\gamma ^\mu\psi\Bigr) u_\mu -\rho\,,\label{Dirpress01}\\
\Pi _\alpha{}^\beta &=& {i\over 2}\Bigl[\overline{\psi}\gamma ^\mu \left( e_\alpha\rfloor \underline{D}\,\psi\right)-\left( e_\alpha\rfloor\overline{\underline{D}\psi}\,\right)\gamma ^\mu\psi\Bigr] h_\mu{}^\beta\,,\label{DirsymmS01}\\
p_\alpha &=& {i\over 2}\Bigl[\overline{\psi}\gamma ^\mu \left( e_\alpha\rfloor \underline{D}\,\psi\right)-\left( e_\alpha\rfloor\overline{\underline{D}\psi}\,\right)\gamma ^\mu\psi\Bigr] u_\mu\,,\label{Dirmomdens01}
\end{eqnarray}
where the bulk viscosity in (\ref{Dirrho01}) is given by
\begin{equation}
S_\mu{}^\mu = \Pi _\mu{}^\mu ={i\over 2}\Bigl[\overline{\psi}\gamma ^\mu \left( e_\mu\rfloor \underline{D}\,\psi\right)-\left( e_\mu\rfloor\overline{\underline{D}\psi}\,\right)\gamma ^\mu\psi\Bigr]\,.\label{DirStrace01}
\end{equation}
Let us take into account the matter field equations
\begin{eqnarray}
i\,{}^*\gamma\wedge D\,\psi -{i\over 2}\,D\eta ^\alpha\gamma _\alpha\psi -\,^*m\,\psi &=&0\,,\label{mattfieldeq1}\\
i\,\overline{D\psi}\wedge\,{}^*\gamma +{i\over 2}\,\overline{\psi}\,D\eta ^\alpha\gamma _\alpha -\,^*m\,\overline{\psi}&=&0\,,\label{mattfieldeq2}
\end{eqnarray}
obtained from (\ref{DiracLagrang1}), where $D\eta ^\alpha = \eta ^\alpha{}_\beta\wedge T^\beta$, see (\ref{antisym2form}), (\ref{antisym3form}) and (\ref{torsiondef}). Rewritting (\ref{mattfieldeq1}) and (\ref{mattfieldeq2}) as
\begin{eqnarray}
i\,\gamma ^\mu e_\mu\rfloor\bigl[ D\psi -{1\over 2}\bigl(e_\nu\rfloor T^\nu\bigr)\psi\bigr] +m\,\psi &=&0\,,\label{mattfieldeq3}\\
i\,e_\mu\rfloor\bigl[\,\overline{D\psi}-{1\over 2}\overline{\psi}\bigl(e_\nu\rfloor T^\nu\bigr)\bigr]\gamma ^\mu -m\,\overline{\psi}&=&0\,,\label{mattfieldeq4}
\end{eqnarray}
and combining them, we find
\begin{eqnarray}
0 &=& {i\over 2}\Bigl(\overline{\psi}\gamma ^\mu \,{\cal \L\/}_u\psi -\overline{{\cal \L\/}_u\psi}\,\gamma ^\mu\psi\Bigr) u_\mu -m\,\overline{\psi}\psi\nonumber\\
&&-{i\over 2}\Bigl[\overline{\psi}\gamma ^\mu \left( e_\mu\rfloor \underline{D}\,\psi\right)-\left( e_\mu\rfloor\overline{\underline{D}\psi}\,\right)\gamma ^\mu\psi\Bigr]\,,\label{mattfieldeq7}
\end{eqnarray}
implying, in view of (\ref{Dirrho01}), (\ref{Dirpress01}) and (\ref{DirStrace01}), the vanishing of pressure, so that from (\ref{Dirpress01}) follows
\begin{equation}
\rho = {i\over 2}\Bigl(\overline{\psi}\gamma ^\mu \,{\cal \L\/}_u\psi -\overline{{\cal \L\/}_u\psi}\,\gamma ^\mu\psi\Bigr) u_\mu\,.\label{Dirrho02}
\end{equation}
Eq.(\ref{Strace01}) with (\ref{Dirrho01}) and zero pressure yields
\begin{equation}
\Sigma _\mu{}^\mu = -m\,\overline{\psi}\psi\,,\label{Direnmomtrace}
\end{equation}
providing a first example of the contribution of mass to the energy-momentum tensor trace. As a further example, let us examine massive matter particles of a different kind.

\subsection{Fundamental bosonic matter}

Looking for a generalization of the particular result (\ref{Direnmomtrace}) for massive matter with no vanishing pressure, we consider the Lagrangian density for scalar matter
\begin{equation}
L^{\rm Bos}=-{1\over 2}\,D\phi ^\dagger\wedge {}^*D\phi -{1\over 2}\,{}^*m^2 \phi ^\dagger {}\phi\,.\label{BoseLagrang1}
\end{equation}
From it we derive the energy-momentum 3-form (\ref{definition03b}) to be
\begin{equation}
\Sigma ^{\rm Bos}_\alpha = {1\over 2}\Bigl[ D\phi ^\dagger\wedge (e_\alpha\rfloor {}^*D\phi ) + {}^*D\phi ^\dagger (\,e_\alpha\rfloor D\phi ) - m^2\phi ^\dagger {}\phi\,\eta _\alpha\Bigr]\,,\label{Bosenergymom01}
\end{equation}
whose constitutive elements (\ref{enmom02}) (the energy-momentum tensor turning out to be symmetric) read
\begin{eqnarray}
\rho &=& {1\over 2}\,\Bigl[ m^2\phi ^\dagger {}\phi +{\cal \L\/}_u\phi ^\dagger {\cal \L\/}_u\phi + (e_\mu \rfloor \underline{D}\phi ^\dagger )(e^\mu \rfloor \underline{D}\phi )\Bigr]\,,\nonumber\\
\label{Bosrho01}\\
q_\alpha &=& {1\over 2}\,\Bigl[ e_\alpha\rfloor\bigl( \underline{D}\phi ^\dagger\,{\cal \L\/}_u\phi + {\cal \L\/}_u\phi ^\dagger\underline{D}\phi\bigr)\Bigr]\,,\label{Bosflux01}\\
p &=& {1\over 2}\,\Bigl[ - m^2\phi ^\dagger {}\phi + {\cal \L\/}_u\phi ^\dagger {\cal \L\/}_u\phi - (e_\mu \rfloor \underline{D}\phi ^\dagger )(e^\mu \rfloor \underline{D}\phi )\Bigr]\,,\nonumber\\
\label{Bospress01}\\
S_{\alpha\beta}&=& (e_{(\alpha } \rfloor \underline{D}\phi ^\dagger )(e_{\beta )}\rfloor \underline{D}\phi )\,.\label{BossymmS01}
\end{eqnarray}
One can easily check that
\begin{eqnarray}
\rho + p &=& {\cal \L\/}_u\phi ^\dagger {\cal \L\/}_u\phi\,,\label{scalrel02}\\
\rho - p &=& m^2\phi ^\dagger {}\phi + S_\mu{}^\mu\,,\label{scalrel03}
\end{eqnarray}
being
\begin{equation}
S_\mu{}^\mu = (e_\mu \rfloor \underline{D}\phi ^\dagger )(e^\mu \rfloor \underline{D}\phi )\,.\label{scalrel01}
\end{equation}
The trace (\ref{Strace01}) calculated from (\ref{scalrel02}) and (\ref{scalrel03}) is 
\begin{equation}
\Sigma _\mu{}^\mu = - m^2\phi ^\dagger {}\phi + 2p\,.\label{Bostrace}
\end{equation}
We will take (\ref{Bostrace}) and the previous result (\ref{Direnmomtrace}) as models for a conjecture on macroscopic matter to be enunciated later.

\subsection{Vacuum}

Let us finally mention that, in the absence of matter and radiation, it is notwithstanding possible to interpret the cosmological constant term as belonging to the {\it matter} sector, as the energy-momentum of vacuum, rather than as a contribution to the gravitational sector, as usual in the treatment of gravity \cite{Chow}. That is, one can consider
\begin{equation}
\Sigma ^{\rm vac}_\alpha = -{1\over{2\kappa}}\Lambda \eta _\alpha\,\label{vacenergymom01}
\end{equation}
as a contribution to (\ref{definition03b}). Comparing (\ref{vacenergymom01}) with the energy-momentum tensor (\ref{enmom02}) reduced to the simple symmetric form
\begin{equation}
\Sigma _\alpha{}^\beta= \bigl(\rho +p\bigr)\,u_\alpha u^\beta +p\,\delta _\alpha ^\beta \,,\label{enmom05}
\end{equation}
describing an unviscid fluid (such that $\Pi _\alpha{}^\beta =0$) with $q_\alpha = p_\alpha =0$, it follows
\begin{eqnarray}
\rho &=& {1\over{2\kappa}}\Lambda\,,\label{vacdens01}\\
p &=& -\rho\,.\label{vacpress01}
\end{eqnarray}
We assume that all forms of matter (phenomenological as much as fundamental), including radiation and even vacuum, contribute to the material energy-momentum, constituting the main source of gravity. This universal principle is supposed to hold regardless of wether a Lagrangian support ensuring it is known or not.

\section{Thermodynamic approach to phenomenological matter}

Having established (\ref{mattender03}) as the general form of the principle of conservation of energy for matter systems of any kind, a suitable interpretation of its variables, together with a few complementary hypotheses, makes this equation able to describe the energetic behavior of macroscopic matter including thermal phenomena. Let us restrict our attention to the thermodynamics of fluids \cite{Landau:1958}. We postulate the energy density $\rho$ in (\ref{mattender03}) to consist of the sum of mass density $\rho _{\rm m}$ and internal (thermal and chemical) energy density $\mathsf{u}$, that is
\begin{equation}
\rho = \rho _{\rm m} + \mathsf{u}\,,\label{rhodecomp01}
\end{equation}
and we introduce a mass current 3-form
\begin{equation}
\epsilon _{\rm m} := -d\tau\wedge\mathsf{J}_{\rm m}^\beta\overline{\eta}_\beta + \rho _{\rm m}\overline{\eta}\,,\label{massflow01}
\end{equation}
including the mass flux contribution $\mathsf{J}_{\rm m}^\beta\overline{\eta}_\beta$. In order to take into account possible nuclear reactions using mass for energy production, we postulate (\ref{massflow01}) to be not necessarily conserved, but in general
\begin{equation}
d\epsilon _{\rm m} \equiv d\tau\wedge\Bigl[ {\it l}_u ( \rho _{\rm m}\overline{\eta}) +\underline{d}\,(\mathsf{J}_{\rm m}^\beta\overline{\eta}_\beta )\Bigr] =-\Theta _{\rm m} \eta\,.\label{massflow02}
\end{equation}
In regard of diffusion and chemical processes \cite{Van:2013sma} \cite{Kremer:2012nk}, we also define particle number current 3-forms built from particle number fourvectors $N_i^\alpha$ for different particle species as
\begin{equation}
N_i^\alpha \eta _\alpha := -d\tau\wedge J_i^\alpha\overline{\eta}_\alpha + n_i\overline{\eta}\,,\label{partfluxdecomp02}
\end{equation}
where $n_i\overline{\eta}$ expresses the average number of particles of a given species contained in the spatial volume element $\overline{\eta}$ (maybe chemical constituents diluted in an inert fluid medium), and $J_i^\alpha\overline{\eta}_\alpha$ is the particle diffusion flux. The particle number current (\ref{partfluxdecomp02}) is assumed not to be in general conserved, even if for ordinary material particles baryon number conservation is in order. The local balance equation for the transfer of different kinds of particles of a mixture into or out of an elementary volume element then reads
\begin{equation}
d\bigl( N_i^\alpha \eta _\alpha \bigr) \equiv d\tau\wedge\Bigl[ {\it l}_u ( n_i\overline{\eta}) +\underline{d}\,(J_i^\alpha\overline{\eta}_\alpha )\Bigr] = -\Theta _i^{_{\rm N}}\eta\,,\label{partfluxdecomp03}
\end{equation}
where $\Theta _i^{_{\rm N}}$ is a source or sink representing the rate of creation or destruction of the number of particles of a given species (due for instance to chemical reactions creating or destroying particles of that species). Thus, according to the continuity equations (\ref{massflow02}) and (\ref{partfluxdecomp03}) for mass and particle number respectively, nuclear and chemical reactions involve in general irreversible fuel consumption. The link of (\ref{massflow01}) and (\ref{partfluxdecomp02}) to the formalism previously developed in Sect.V is established by postulating the energy flux (\ref{eneryflux01}) to consist of the sum of heat, mass and chemical flows as
\begin{equation}
q^\beta = q_{_H}^\beta -\mathsf{J}_{\rm m}^\beta -\mu ^i J_i^\beta\,,\label{energfluxdecomp01}
\end{equation}
where we introduced the chemical potentials $\mu ^i$ as additional quantities. In the present approach, it is superfluous to consider an entropy flux \cite{DeGroot:1980dk} \cite{Van:2017} neither as an independent quantity nor as defined in terms of the flows $q_{_H}^\beta$ and $\mu ^i J_i^\beta$ appearing in (\ref{energfluxdecomp01}).

\subsection{The first law of thermodynamics}

The gauge theoretical theorem of conservation of energy (\ref{mattender03}) makes contact with the phenomenology of thermal processes by substituting (\ref{rhodecomp01}) and (\ref{energfluxdecomp01}) into (\ref{mattender03}) taking into account (\ref{massflow02}) and (\ref{partfluxdecomp03}), so that one gets
\begin{eqnarray}
{\it l}_u (\mathsf{u}\overline{\eta}) &-& \underline{d}\,(q_{_H}^\beta\overline{\eta}_\beta ) + p\,{\it l}_u \overline{\eta} -\mu ^i {\it l}_u ( n_i\overline{\eta})\nonumber\\
&+& {\cal \L\/}_u \underline{\vartheta}^\alpha\wedge\Pi _\alpha{}^\beta\overline{\eta}_\beta + \underline{d}\mu ^i\wedge J_i^\beta\overline{\eta}_\beta\nonumber\\
&-& p_\alpha\,{\cal \L\/}_u u^\alpha\,\overline{\eta} -T_{\bot}^\alpha\,u_\alpha \wedge q^\beta\overline{\eta}_\beta\nonumber\\
&-& R_{\bot}^{\alpha\beta}\wedge{(\tau _{\alpha\beta})}_{\bot} -\mu ^i\Theta _i^{_{\rm N}}\overline{\eta} -\Theta _{\rm m}\overline{\eta} =0\,.\nonumber\\
\label{modiffirstlaw01}
\end{eqnarray}
Eq. (\ref{modiffirstlaw01}) becomes more familiar when written in a suitable notation. At this purpose we introduce the following definitions
\begin{eqnarray}
\mathfrak{U} &:=& \mathsf{u}\overline{\eta}\,,\label{intenergy01}\\
\mathfrak{q}_{_H} &:=& q_{_H}^\beta\overline{\eta}_\beta\,,\label{heatflux01}\\
\mathcal{N}_i &:=& n_i\overline{\eta}\,,\label{partnumb01}\\
\Theta _{\rm tot}\overline{\eta} &:=& -\bigl( e_\beta\rfloor {\cal \L\/}_u \underline{\vartheta}^\alpha\bigr)\Pi _\alpha{}^\beta\overline{\eta} -\bigl( e_\beta\rfloor\underline{d}\mu ^i\bigr) J_i^\beta\overline{\eta}\nonumber\\
&&+ p_\alpha\,{\cal \L\/}_u u^\alpha\,\overline{\eta}+T_{\bot}^\alpha\,u_\alpha \wedge q^\beta\overline{\eta}_\beta\nonumber\\
&&+ R_{\bot}^{\alpha\beta}\wedge{(\tau _{\alpha\beta})}_{\bot} +\mu ^i\Theta _i^{_{\rm N}}\overline{\eta} +\Theta _{\rm m}\overline{\eta}\,,\label{Thetatot01}
\end{eqnarray}
being (\ref{intenergy01}) the amount of internal energy stored in an elementary volume, (\ref{heatflux01}) the heat flux, (\ref{partnumb01}) the number of particles of a given species contained in the volume element considered, and (\ref{Thetatot01}) the energy dissipation rate due to irreversible processes (as interpreted from (\ref{secondlawthermsimpl}) below). In terms of these quantities, (\ref{modiffirstlaw01}) takes the form
\begin{equation}
{\it l}_u\,\mathfrak{U} -\underline{d}\mathfrak{q}_{_H} + p\,{\it l}_u \overline{\eta} -\mu ^i\,{\it l}_u \mathcal{N}_i = \Theta _{\rm tot}\overline{\eta}\,,\label{firstlawtherm01}
\end{equation}
in which one easily recognizes the first law of thermodynamics as a local energy balance equation for thermal phenomena. Eq.(\ref{firstlawtherm01}) establishes the conservation of energy as a multiform quantity subject to various transfer and conversion processes taking place between a given thermodynamic system and its surroundings. When one integrates (\ref{firstlawtherm01}) in a closed domain ${\cal D}$ of 3-dimensional space enclosed by the boundary $\partial {\cal D}$, the rate of change of the amount of internal energy accumulated in the volume ${\cal D}$ is shown to depend on the energy interchange with the exterior, measured by the balance of inflow and outflow of the heat flux passing across the enclosure $\partial {\cal D}$ plus the work performed by or done on external bodies by energy transfer, energetic changes involved in chemical reactions and further contributions to be interpreted later, in view of the second law, as dissipation (heating) due irreversible processes. The interconvertibility between the different forms of energy means that, in order to preserve energy from creation nor destruction, work production requires either heat supply to the system or internal energy consumption, the latter allowing to produce work in adiabatic processes at no expense of any external heat source. Conversely, work done on the system, or incoming heat not spent in doing work, increase the amount of internal energy stored. In general, heat passing through the system is partially absorbed and partially transformed into work. Observable effects of energy transfer such as heat absorption or work generation are external phenomena occurring outside the system's boundary, the latter playing a fundamental role in thermodynamics. Certain energetic and material exchanges may or may not be allowed according to wether the enclosing surface of each small portion of the global system is rigid or movable, penetrable or impenetrable to matter, or consists either of adiabatic or diathermal walls. In particular, open systems can swap energy and matter, closed systems only energy, and isolated systems none of both. In our approach, no restrictions to any kind of interchange are considered in principle.

The energy conservation law (\ref{firstlawtherm01}) does not set any restriction on the possible conversions between different forms of energy. This total ability to mutate remains in force as far as transformations among internal energy and work, or of any of them into heat, are concerned (as for instance in the case of total conversion of work into heat by heating a body by friction). However, the conversion of heat into other forms of energy is severely limited by the second law.

\subsection{The second law of thermodynamics}

The present (gauge) treatment of irreversible processes, to be completed next, is closely related to {\it classical irreversible thermodynamics}, pioneered by Onsager and Prigogine \cite{Nicolis:1939} \cite{Kondepudi:1952}, resting on the {\it local equilibrium hypothesis}, according to which thermodynamic quantities are locally well defined in inhomogeneous systems out of equilibrium, being subject instantaneously at each point to the same relations as assumed globally for uniform systems in equilibrium in ordinary thermodynamics \cite{Nicolis:1939} -\cite{Lebon:2008}. Macroscopic matter systems are supposed to decompose into infinitesimal subsystems endowed with dynamical and thermal properties described by spacetime dependent variables fulfilling local thermodynamical laws. In this spirit, we assume the internal energy $\mathfrak{U}$ to be a functional of macroscopic physical quantities, analogous to the standard thermodynamical ones, represented in our case by differential forms. In particular, for the description of fluids, we choose
\begin{equation}
\mathfrak{U} = \mathfrak{U}\,(\mathfrak{s}\,,\mathcal{N}_i\,,\overline{\eta}\,)\,,\label{uargs}
\end{equation}
being $\overline{\eta}$ an infinitesimal volume element, $\mathcal{N}_i := n_i\overline{\eta}$ the number of particles of various species --from subatomic to molecular-- enclosed in it, and the 3-form $\mathfrak{s} =\sigma \overline{\eta}$ the entropy, as a new physical variable necessary to deal with macroscopic material systems \cite{Callen}. (Densities represent extensive thermodynamical variables.) The Lie derivative of the internal energy (\ref{uargs}) expands as
\begin{equation}
{\it l}_u\,\mathfrak{U} = {{\partial\mathfrak{U}}\over{\partial\mathfrak{s}}}\,{\it l}_u\mathfrak{s}
+{{\partial\mathfrak{U}}\over{\partial \mathcal{N}_i}}\,{\it l}_u \mathcal{N}_i
+{{\partial\mathfrak{U}}\over{\partial \overline{\eta}}}\,{\it l}_u \overline{\eta}\,,\label{uLiederiv01}
\end{equation}
where the functional derivatives are defined \cite{Callen} as
\begin{equation}
{{\partial\mathfrak{U}}\over{\partial\mathfrak{s}}} =: T\,,\quad
{{\partial\mathfrak{U}}\over{\partial \mathcal{N}_i}} =: \mu ^i\,,\quad
{{\partial\mathfrak{U}}\over{\partial \overline{\eta}}} =: -p\,,\label{Uders01}
\end{equation}
That is, the derivative of ${\mathfrak{U}}$ with respect to the entropy is identified as absolute temperature (absent from (\ref{firstlawtherm01})), its rate of change with respect to the number of particles of each species is measured by a chemical potential (the potential energy involved in chemical reactions or phase transitions), and the derivative of ${\mathfrak{U}}$ with respect to the volume defines pressure. Thus, (\ref{uLiederiv01}) with (\ref{Uders01}) reads
\begin{equation}
{\it l}_u\,\mathfrak{U} = T\,{\it l}_u\mathfrak{s} +\mu ^i\,{\it l}_u \mathcal{N}_i -p\,{\it l}_u \overline{\eta}\,,\label{uLiederiv02}
\end{equation}
which is the Gibbs fundamental equation for internal energy. The term $T\,{\it l}_u{\mathfrak s}$ measures the quantity of heat interchanged with the surroundings at a given temperature. In particular, heat received by a system can be used to increase its internal energy (heating), to do mechanical work (volume expansion) or to perform changes of phases or of chemical composition (consisting, roughly speaking, in adding or substracting particles). The compatibility requirement between the first law of thermodynamics (\ref{firstlawtherm01}) deduced above and the alternative formulation (\ref{uLiederiv02}) of the same physical principle gives rise to the second law of thermodynamics
\begin{equation}
{\it l}_u\mathfrak{s} - {1\over T}\,\underline{d}\mathfrak{q}_{_H} = {1\over T}\,\Theta _{\rm tot}\overline{\eta}\,,\label{secondlawthermsimpl}
\end{equation}
providing an additional condition on the entropy ignored in (\ref{uLiederiv02}), where it was introduced for the first time. According to (\ref{secondlawthermsimpl}), the entropy production or consumption in a system originates exclusively from heat exchange or energy dissipation. Contrarily to changes of internal energy, entropy increase cannot be induced by doing work on the system. Heat supply and absorption also differs from other forms of energy transfer in that it requires the concurrence of surrounding bodies at different temperatures. Being the concept of temperature absent from the first law in its form (\ref{firstlawtherm01}), it is the second law (\ref{secondlawthermsimpl}) that, in terms of the mutually related notions of temperature and entropy, determines the directionality of spontaneous heat transfer from hotter to colder bodies, provided
\begin{equation}
\Theta _{\rm tot}\geq 0\,.\label{nonnegtheta01}
\end{equation}
This non-negativity condition on the source term in (\ref{secondlawthermsimpl}) --measuring there the total entropy generated by irreversible processes-- does in general not follow necessarily from the present approach, but depends on the particular matter model considered. Requirement (\ref{nonnegtheta01}) is to be regarded as an externally imposed criterion, based on experience, for the physical acceptability of realistic models.

As an illustration of possible conditions ensuring (\ref{nonnegtheta01}) to hold, the first term in the rhs of (\ref{Thetatot01}) results to be non-negative if we consider a generalization of standard Newtonian fluids giving rise to a sum of quadratic terms as follows. From (\ref{Stresstens01})-(\ref{tracefreeS01}) we find that the total stress tensor decomposes as
\begin{equation}
\Pi _{\alpha\beta} =\Pi _{[\alpha\beta ]} + {\slash\!\!\!\!S}_{\alpha\beta} +{1\over 3} h_{\alpha\beta} S_\mu{}^\mu\,.\label{pidecomp}
\end{equation}
The contributions of shear and bulk viscosities are both positive (or zero) provided one postulates for them to be
\begin{eqnarray}
{\slash\!\!\!\!S}_{\alpha\beta} &=& -2\eta \Bigl[ e_{(\alpha}\rfloor {\cal \L\/}_u \underline{\vartheta}_{\beta )} -{1\over 3}\,h_{\alpha\beta}\bigl( e_\mu\rfloor {\cal \L\/}_u \underline{\vartheta}^\mu\bigr)\Bigr]\,,\label{positcond02}\\
S_\mu{}^\mu &=& -3\zeta\,( e_\mu\rfloor {\cal \L\/}_u \underline{\vartheta}^\mu\bigr)\,,\label{positcond03}
\end{eqnarray}
respectively, built in term of ${\cal \L\/}_u\underline{\vartheta}^\alpha$ as a generalized velocity gradient, see (\ref{thetaLiederiv04}). Analogously, the antisymmetric piece in (\ref{pidecomp}) may be taken as
\begin{equation}
\Pi _{[\alpha\beta ]} = \gamma \bigl( e_{[\alpha}\rfloor {\cal \L\/}_u \underline{\vartheta}_{\beta ]}\bigr)\,.\label{positcond01}
\end{equation}
For the second term in the rhs of (\ref{Thetatot01}), it is possible to assume the chemical diffusion law
\begin{equation}
J_i^\beta = -\textsc{L}_i{}^j \bigl( e^\beta\rfloor \underline{d}\mu _j\bigr)\,,\label{positcond04}
\end{equation}
more fundamental than Fick's empirical first law \cite{Nicolis:1939} \cite{Brady}, where the particle diffusion flux is generated by gradients in the chemical potentials, and $\textsc{L}_i{}^j$ is a positive definite matrix of phenomenological diffusion coefficients. The non-negativity of the remaining terms is more problematic. Regarding stars, the nonnegative contribution of $\Theta _{\rm m}$ is granted to some extent since the decrease of the amount of mass due to fusion gives rise to positive heat production. But $\mu ^i\Theta _i^{_{\rm N}}$ is positive only for exothermal chemical reactions and the sign of $p_\beta\,{\cal \L\/}_u u^\beta$ is undetermined in principle, so that one cannot but merely assume it, if negative, to be no greater than the sum of the other contributions. The same holds for the term involving torsion. For the spin term, one could make the Ohm's law-like assumption
\begin{eqnarray}
\tau _{\bot}^{\alpha\beta} = \xi\,^\# R_{\bot}^{\alpha\beta}\,.\label{positcond05}
\end{eqnarray}
In any case, the fulfillment of (\ref{nonnegtheta01}) is not {\it a priori} guaranteed for arbitrary kinds of matter and it remains an open question to determine which ones are compatible with that key thermodynamical requirement.

Regarding other main aspect of the second law (\ref{secondlawthermsimpl}), the quantity of heat flow $\mathfrak{q}_{_H}$ penetrating the system across its bounding surface is weighted by the temperature of the external body involved at each stage of the transfer process. The temperatures of the system and of the outer heat sources are in general distinct from each other even though, eventually, in reversible processes or during transient local equilibrium states, the temperature difference may tend to zero. Heat supply ${{\underline{d}\mathfrak{q}_{_H}}\over T}$ from the exterior, along with entropy production ${1\over T}\,\Theta _{\rm tot}\overline{\eta}$, related to energy dissipation, give rise to a change ${\it l}_u\mathfrak{s}$ over time in the amount of entropy stored in the system. In order for heat transfer to take place spontaneously, it must be accompanied by a global growth of entropy in the joint system formed by both, the studied system and its environment. Regarded locally, this means that, whenever an entropy decrease is caused by a small system in its surroundings by taking heat from an external thermal source at a given temperature, the system should compensate the loss by giving up at least part of the absorbed heat to an external sink at a different temperature -since no work but only heat exchanges produce entropy alterations--, so that the amount of entropy surrendered by the local system is larger than the one it received, and, on balance, a global increase of entropy occurs. In reversible thermodynamic cycles (in which the system returns back to its initial state while $\Theta _{\rm tot}= 0$), incoming heat at a higher temperature is balanced by a smaller quantity of outgoing heat at a lower temperature so that $\oint{{\underline{d}\mathfrak{q}_{_H}}\over T} =0$. Provided the system interchanges heat successively with a source and a sink at constant temperatures $T_{\rm in}$ and  $T_{\rm out}$ respectively, integration of the latter condition yields
\begin{equation}
{{Q_{\rm out}}\over{T_{\rm out}}} = {{Q_{\rm in}}\over{T_{\rm in}}}\,,\label{cyclheatexch02}
\end{equation}
being $Q_{\rm in}- Q_{\rm out } = Q_{\rm in}\bigl( 1 -{{T_{\rm out}}\over{T_{\rm in}}}\bigr)$ the amount of heat an engine working between the given temperatures can use to produce work.

Recall that the two laws constituting the general unifying principles for all physical phenomena involving heat interchanges are deduced --up to the phenomenological condition (\ref{nonnegtheta01})-- from local translational invariance as considered in PGT, thus deserving to be dubbed laws of {\it gauge thermodynamics} as proposed in the title of the present paper.

\section{State equations and thermodynamical behavior of simple no spin fluids}

Thermodynamics provides the abstract framework to deal with energy, but it must be completed with empirical state equations concerning concrete physical quantities. On the theoretical site, the Gibbs fundamental law (\ref{uLiederiv02}) with (\ref{intenergy01}), (\ref{partnumb01}) and $\mathfrak{s} =\sigma \overline{\eta}$ expands as
\begin{equation}
{\it l}_u\mathsf{u}\,\overline{\eta} + \mathsf{u}\,{\it l}_u\overline{\eta} = \bigl( T{\it l}_u\sigma +\mu ^i{\it l}_u n_i\bigr)\overline{\eta} + \bigl( T\sigma + \mu ^i n_i -p\bigr){\it l}_u\overline{\eta}\,.\label{uLiederiv03}
\end{equation}
The simplest way to ensure (\ref{uLiederiv03}) to hold is to assume the separate conditions
\begin{eqnarray}
{\it l}_u\mathsf{u} &=& T{\it l}_u\sigma +\mu ^i{\it l}_u n_i\,,\label{ansatz01}\\
\mathsf{u} + p &=& T\sigma +\mu ^i n_i\,,\label{ansatz02}
\end{eqnarray}
being the quantity in the lhs of (\ref{ansatz02}) the enthalpy density. The link between theory and phenomenology is established by requiring (\ref{ansatz01}) and (\ref{ansatz02}) to be satisfied by the variables related to each other by the state equations supported by observation. This provides us with precise values of the quantities $\mathsf{u}$, $p$, $\sigma$ and $\mu ^i$, from which we further will require to fit consistently in the general scheme determined by energy-momentum conservation. In Sect.IX we will derive the consequences of this requirement for energy flux and shear viscosity.

\subsection{Ideal monatomic gases}

Let us first consider the phenomenological equation of state of perfect gases, summarizing the laws of Boyle-Mariotte, Charles and Gay-Lussac, and Avogadro. It establishes the simple relation $pV=n_{\rm mol}RT$ between pressure, volume and temperature, being $n_{\rm mol}$ the number of moles and $R$ the ideal gas constant. Although not strictly necessary, in order to reexpress the same state equation in terms of the particle number density present in (\ref{ansatz01}) and (\ref{ansatz02}), we take from statistical mechanics \cite{Hill:2011} a result according to which the laws of atomic and subatomic processes involving discrete particles predict the observable behavior of macroscopic {\it continuous} matter. By considering the classical Maxwell-Boltzmann statistical approach to ideal dilute monatomic gases constituted by point masses of particles of a single species without structure and with no mutual interactions other than perfectly elastic collisions, the state equation takes the form $p = k_{_B}nT$, where $k_{_B}$ is the Boltzmann constant and $n$ the particle number density. The remaining relevant thermodynamic quantities fulfilling (\ref{ansatz01}) and (\ref{ansatz02}) can be also derived, resulting to be
\begin{eqnarray}
\mathsf{u} &=& {3\over 2}\,k_{_B} n T\,,\label{idgasenergydens02}\\
p &=& k_{_B} n T\,,\label{idgaspress02}\\
\sigma &=& k_{_B} n \Biggl( \log{\Biggl[ {{\bigl( k_{_B} T\bigr)^{3\over 2}}\over n}\Biggr]}+{5\over 2}+\alpha _{_0}\,\Biggr)\,,\label{idgasentropy02}\\
\mu &=& -k_{_B} T \Biggl( \log{\Biggl[ {{\bigl( k_{_B} T\bigr)^{3\over 2}}\over n}\Biggr]}+\alpha _{_0}\,\Biggr)\,.\label{idgaschempot02}
\end{eqnarray}
The constant $\alpha _{_0}$, and thus the absolute value of entropy, left undetermined by the classical approach, can be calculated with the help of quantum mechanics. The Sackur-Tetrode equation for entropy \cite{Grimus:2011} fixes it as $\alpha _{_0}= {3\over 2}\log{\Bigl({{2\pi m}\over{h^2}}\Bigr)}$, being $h$ Planck's constant and $m$ the mass of a single particle. The quantum ideal gas results, obtained in the limit of low densities and high temperatures, coincide with those (\ref{idgasenergydens02})-(\ref{idgaschempot02}) of the classical approach, regardless of whether Bose-Einstein or Fermi-Dirac statistics is used to deduce them \cite{Goodstein:1985}. No matter in what way one arrives at Eqs.(\ref{idgasenergydens02})-(\ref{idgaschempot02}), the relevant fact is that they provide a set of physically meaningful quantities approaching the observational behavior of gases as reflected in the state equation (\ref{idgaspress02}) and satisfying at the same time (\ref{ansatz01}) and (\ref{ansatz02}) particularized to the case of a gas constituted by a single species of particles.

\subsection{Van der Waals fluids}

A generalization of the previous result is provided by the Van der Waals state equation
\begin{equation}
\bigl( p + a n^2\bigr)\bigl( 1 -b n\bigr) = k_{_B} nT\,,\label{VdWgaspress01}
\end{equation}
where interactions between particles are taken into account \cite{Hill:2011}. In immediate analogy to (\ref{idgasenergydens02})-(\ref{idgaschempot02}) (as the simplest particular case), we use (\ref{ansatz01}) and (\ref{ansatz02}) to derive the quantities
\begin{eqnarray}
\mathsf{u} &=& c_{\rm n} k_{_B} n T - a n^2\,,\label{VdWgasenergydens02}\\
p &=& {{k_{_B} n T}\over{\bigl( 1 -b n\bigr)}}-a n^2\,,\label{VdWgaspress02}\\
\sigma  &=& k_{_B} n\Biggl\{ \log{\Biggl[ {{\bigl( k_{_B} T\bigr)^{c_{\rm n}}}\over n}\bigl(1 -b n\bigr)\Biggr]} + \bigl( c_{\rm n} +1\bigr) + A_{_0}\,\Biggr\}\,,\nonumber\\
\label{VdWgasentropy02}\\
\mu &=& -k_{_B} T\Biggl\{ \log{\Biggl[ {{\bigl( k_{_B} T\bigr)^{c_{\rm n}}}\over n}\bigl(1 -b n\bigr)\Biggr]} -{{b n}\over{\bigl( 1 -b n\bigr)}}
+ A_{_0}\,\Biggr\}\nonumber\\
&& - 2 a n \,,\label{VdWgaschempot02}
\end{eqnarray}
where we introduced 
\begin{equation}
c_{\rm n}:= {1\over{k_{_B} n}}\Bigl( {{\partial\mathsf{u}}\over{\partial T}}\Bigr)_{\rm n}\label{specheat02}
\end{equation}
as the redefined specific heat capacity at constant number density. As before, (\ref{VdWgasenergydens02})-(\ref{VdWgaschempot02}) fulfill both, the equation of state (\ref{VdWgaspress01}) (that is, (\ref{VdWgaspress02})), and the Gibbs conditions (\ref{ansatz01}) and (\ref{ansatz02}) restricted to a single species of particles as
\begin{eqnarray}
{\it l}_u\mathsf{u} &=& \mu {\it l}_u n + T{\it l}_u\sigma\,,\label{ansatz03}\\
\mathsf{u} + p &=& T\sigma +\mu n\,.\label{ansatz04}
\end{eqnarray}
The functional dependence of (\ref{VdWgasenergydens02}) on temperature and particle number density has as a consequence that, in general, changes in the internal energy of a system induced by energy transfer manifest themselves as heating or cooling, but they may also be due to changes in the particle number density resulting from diffusion processes or chemical reactions.

\subsection{Photon gas}

Thermal radiation of a body in thermal equilibrium is described using as a theoretical model a photon gas filling a cavity \cite{Prigogine} \cite{Demirel}. All physical quantities involved in it are functions of the temperature
\begin{eqnarray}
\mathsf{u}_{\rm ph} &=& \alpha T^4\,,\label{photgasenergydens01}\\
p_{\rm ph} &=& {1\over 3}\,\mathsf{u}_{\rm ph}\,,\label{photgaspress01}\\
\sigma _{\rm ph}  &=& {4\over 3}\,\alpha T^3\,,\label{photgasentropy01}\\
\mu _{\rm ph} &=& 0\,,\label{photgaschempot01}\\
n_{\rm ph} &=& \lambda T^3\,,\label{photnumb01}
\end{eqnarray}
automatically satisfying (\ref{ansatz03}) and (\ref{ansatz04}). Notice that (\ref{photgaspress01}) reproduces the relation (\ref{emeffpress01}) between pressure and energy density previously found for Maxwell fields. In the following, we will assume (\ref{photgasenergydens01})-(\ref{photnumb01}) to hold even in irreversible processes.

\section{Spherical matter sources of gravity}

Next we consider matter as source of gravity. Although, in general, the sources of the gravitational fields involved in PGT are the energy-momentum tensor and the spin current, we will restrict our attention to the simple case of no-spin fluids. Let us consider a Lagrangian density (\ref{totalLag}) consisting of the sum of a non-specified matter piece and a gravitational contribution given by the Hilbert-Einstein term
\begin{eqnarray}
L^{\rm gr}=-{1\over{2\kappa}}\,\,\eta_{\alpha\beta}\wedge R^{\alpha\beta}\,,\label{gravlagr}
\end{eqnarray}
with $\kappa =8\pi G$ as the gravitational constant, being $G$ Newton's constant. The features of matter will be determined later with the help of the field equations and the conservation laws. From (\ref{gravlagr}) we find the gravitational energy-momentum 3-form (\ref{defingrem}) to be
\begin{equation}
E_\alpha = -{1\over{2\kappa}}\,\eta _{\alpha\mu\nu}\wedge R^{\mu\nu}\,,\label{Einstgravmom}
\end{equation}
that is, the exterior calculus formulation of the Einstein tensor, while the translational and Lorentz excitations (\ref{definition02}) read respectively
\begin{eqnarray}
H_\alpha &=& 0\,,\label{torsmom}\\
H_{\alpha\beta}&=&{1\over{2\kappa}}\,\eta_{\alpha\beta}\,.\label{curvmom}
\end{eqnarray}
In terms of these particular values, the field equations (\ref{fieldeq1}) and (\ref{fieldeq2}) reduce to
\begin{eqnarray}
{1\over{2\kappa}}\,\eta _{\alpha\mu\nu}\wedge R^{\mu\nu} &=& \Sigma _\alpha\,,\label{fieldeq1red}\\
{1\over{2\kappa}}\,D\eta_{\alpha\beta} &=& 0\,,\label{fieldeq2red}
\end{eqnarray}
provided we assume, as already announced, the (non specified) matter Lagrangian to describe a fluid with zero spin current (\ref{definition03c}). In this case, (\ref{spincurrconserv}) reduces to
\begin{equation}
\vartheta _{[\alpha}\wedge \Sigma _{\beta ]} = 0\,,\label{symmsigma01}
\end{equation}
establishing that the matter energy-momentum must be symmetric. On the other hand, from (\ref{fieldeq2red}), taking into account that $D\eta_{\alpha\beta} = \eta_{\alpha\beta\gamma}\wedge T^\gamma$, follows the vanishing of torsion, so that (\ref{fieldeq1red}) becomes the standard Einstein equation we are going to solve.

Looking for an inner Schwarzschild solution for gravity originated by a non-rotating spherical distribution of no-spin matter, we postulate the coframe
\begin{eqnarray}
\vartheta ^0 &=& f dt\,,\label{cofr01}\\
\vartheta ^1 &=& {{dr}\over g}\,,\label{cofr02}\\
\vartheta ^2 &=& r\,d\theta\,,\label{cofr03}\\
\vartheta ^3 &=& r\sin{\theta}\,d\varphi\,,\label{cofr04}
\end{eqnarray}
constituted by spherically symmetric tetrads with functions $f=f(t,r)$ and $g=g(t,r)$ of the radial and temporal variables. The non-vanishing components of the Christoffel connections
\begin{equation}
\Gamma _{\alpha\beta} = e_{[\alpha }\rfloor d\vartheta _{\beta ]} -{1\over 2}\,( e_\alpha\rfloor e_\beta\rfloor d\vartheta _\gamma )\,\vartheta ^\gamma \,,\label{Christoffdef}
\end{equation}
calculated from the tetrads (\ref{cofr01})-(\ref{cofr04}), read
\begin{eqnarray}
\Gamma _{01} &=& \Bigl(\,g\,\partial _r\log f\Bigr)\vartheta ^0 -\Bigl({1\over f}\,\partial _t\log g\Bigr)\vartheta ^1\nonumber\\
&=& (e_1\rfloor d\log{f})\,\vartheta ^0 - (e_0\rfloor d\log{g})\,\vartheta ^1\,,\label{Christoff01}\\
\Gamma _{02} &=& 0\,,\label{Christoff02}\\
\Gamma _{03} &=& 0\,,\label{Christoff03}\\
\Gamma _{12} &=& {g\over r}\,\vartheta ^2\,,\label{Christoff04}\\
\Gamma _{13} &=& {g\over r}\,\vartheta ^3\,,\label{Christoff05}\\
\Gamma _{23} &=& {{\cos{\theta}}\over{r\sin{\theta}}}\,\vartheta ^3\,,\label{Christoff06}
\end{eqnarray}
where --in (\ref{Christoff01})-- we used the dual frame of (\ref{cofr01})-(\ref{cofr04}), that is
\begin{eqnarray}
e_0 &=& {1\over f}\,\partial _t\,,\label{fr01}\\
e_1 &=& g\,\partial _r\,,\label{fr02}\\
e_2 &=& {1\over r}\,\partial _\theta\,,\label{fr03}\\
e_3 &=& {1\over{r\sin{\theta}}}\,\partial _\varphi\,.\label{fr04}
\end{eqnarray}
The non-zero components of the curvature
\begin{equation}
R_\alpha{}^\beta := d\,\Gamma _\alpha{}^\beta + \Gamma _\gamma{}^\beta\wedge \Gamma _\alpha{}^\gamma\,\label{curvdef}
\end{equation}
calculated from (\ref{Christoff01})-(\ref{Christoff06}) are found to be
\begin{eqnarray}
R^{01} &=&  {g\over f}\Bigl[ \partial _r (g\,\partial _r f)-\partial _t \Bigl( {1\over f}\partial _t\Bigl( {1\over g}\Bigr)\Bigr)\Bigl]\vartheta ^0\wedge\vartheta ^1\,,\label{curv01}\\
R^{02} &=& {{g^2}\over{r}}\partial _r\log f\,\vartheta ^0\wedge\vartheta ^2  - {{1}\over{fr}}\partial _t g\,\vartheta ^1\wedge\vartheta ^2\,,\label{curv02}\\
R^{03} &=& {{g^2}\over{r}}\partial _r\log f\,\vartheta ^0\wedge\vartheta ^3  - {{1}\over{fr}}\partial _t g\,\vartheta ^1\wedge\vartheta ^3\,,\label{curv03}\\
R^{12} &=& {{1}\over{fr}}\partial _t g\,\vartheta ^0\wedge\vartheta ^2  + {1\over{r}}g\,\partial _r g\,\vartheta ^1\wedge\vartheta ^2\,,\label{curv04}\\
R^{13} &=& {{1}\over{fr}}\partial _t g\,\vartheta ^0\wedge\vartheta ^3  + {1\over{r}}g\,\partial _r g\,\vartheta ^1\wedge\vartheta ^3\,,\label{curv05}\\
R^{23} &=& {{(g^2-1)}\over{r^2}}\,\vartheta ^2\wedge\vartheta ^3\,.\label{curv06}
\end{eqnarray}
Substituting (\ref{curv01})-(\ref{curv06}) into (\ref{fieldeq1red}), we get, for the non-vanishing tensorial components of the matter energy-momentum (\ref{enmom01}), the conditions
\begin{eqnarray}
\kappa\,\Sigma _0{}^0  &=& {1\over{r^2}}\partial _r \bigl[ r ( g^2 -1)\bigr] \,,\label{Einsteq01}\\
\kappa\,\Sigma _1{}^0  &=& {{2}\over{fr}}\partial _t g \,,\label{Einsteq02}\\
\kappa\,\Sigma _1{}^1  &=& 2\,{{g^2}\over{r}}\partial _r\log f +{{(g^2-1)}\over{r^2}}\,,\label{Einsteq03}\\
\kappa\,\Sigma _2{}^2  &=& {g\over f}\Bigl[ \partial _r (g\,\partial _r f)-\partial _t \Bigl( {1\over f}\partial _t\Bigl( {1\over g}\Bigr)\Bigr)\Bigl]\nonumber\\
&&+{{g^2}\over{r}}\partial _r\log f +{1\over{r}}g\,\partial _r g\,,\label{Einsteq04}\\
\kappa\,\Sigma _3{}^3  &=& \kappa\,\Sigma _2{}^2\,.\label{Einsteq05}
\end{eqnarray}
From (\ref{Einsteq01}) follows
\begin{equation}
g^2 =1 -{{\kappa}\over{8\pi}}{{2m}\over{r}}\,,\label{gfunct01}
\end{equation}
in terms of the mass function $m = m(t,r)$ defined as
\begin{equation}
m:= -\int _0^r\Sigma _0{}^0\,4\pi r^2 dr\,,\label{gfunct02}
\end{equation}
so that
\begin{equation}
\Sigma _0{}^0 =-{1\over{4\pi r^2}}\,\partial _r m\,.\label{gfunct03}
\end{equation}
Eq. (\ref{Einsteq02}) with (\ref{gfunct01}) yields
\begin{equation}
\Sigma _1{}^0 = {{2}\over{\kappa fr}}\partial _t g = -{{\partial _t m}\over{4\pi r^2 fg}}\,,\label{gfunct04}
\end{equation}
and, on the other hand, the difference between (\ref{Einsteq01}) and (\ref{Einsteq03}) reads
\begin{equation}
2\,{{g^2}\over{r}}\partial _r\log \Bigl({f\over g}\Bigr) = \kappa\,\bigl( \Sigma _1{}^1 -\Sigma _0{}^0\bigr)\,,\label{ffunct01}
\end{equation}
implying
\begin{equation}
f = g\exp{\Bigl({{\kappa}\over{8\pi}}\chi\Bigr)}\,,\label{ffunct02}
\end{equation}
where the function $\chi (t,r)$ is defined as
\begin{equation}
\chi := \int _0^r\Bigl( \Sigma _1{}^1 -\Sigma _0{}^0\Bigr){1\over{g^2}} 4\pi r dr\,,\label{ffunct03}
\end{equation}
giving rise to
\begin{equation}
\Sigma _1{}^1 -\Sigma _0{}^0 = {{g^2}\over{4\pi r}}{\partial _r\chi }\,.\label{ffunct04}
\end{equation}
Integration functions depending on $t$ are chosen to vanish in order to ensure the matching with the Schwarzschild vacuum solution, for which $f=g$ and $\Sigma _1{}^1 =\Sigma _0{}^0$. Suitably rewriting the energy-momentum components (\ref{gfunct03}), (\ref{gfunct04}) and (\ref{ffunct04}) with the help of (\ref{fr01}) and (\ref{fr02}), and expressing (\ref{Einsteq04}) in terms of them, (\ref{Einsteq01})-(\ref{Einsteq05}) take the form
\begin{eqnarray}
\Sigma _0{}^0 &=& -{1\over{4\pi r^2 g}}\,(e_1\rfloor dm)\,,\label{Minksig00}\\
\Sigma _1{}^0 &=& -{1\over{4\pi r^2 g}}\,(e_0\rfloor dm)\,,\label{Minksig10}\\
\Sigma _1{}^1 &=& \Sigma _0{}^0 + {g\over{4\pi r}}\,(e_1\rfloor d\chi)\,,\label{Minksig11}\\
\Sigma _2{}^2 &=& \Sigma _1{}^1 +{r\over{2g}}\,\Biggl[ e_0\rfloor d\Sigma _1{}^0 + e_1\rfloor d\Sigma _1{}^1\nonumber\\
&&\hskip1.7cm -2\,(e_0\rfloor d\log{g})\Sigma _1{}^0\nonumber\\
&&\hskip1.7cm +(e_1\rfloor d\log{f})\bigl(\Sigma _1{}^1 -\Sigma _0{}^0\bigr)\Biggr]\,,\label{Minksig22}\\
\Sigma _3{}^3  &=& \Sigma _2{}^2\,.\label{Minksig33}
\end{eqnarray}
The derivatives of $f$ and $g$ appearing in (\ref{Minksig22}) are found, from (\ref{gfunct01}) and (\ref{ffunct02}), to be respectively
\begin{eqnarray}
d\log g &=& {{\kappa r}\over{2 g}}\,\Bigl( -{1\over{4\pi r^2 g}}\,dm +{{m}\over{4\pi r^3}}\,\vartheta ^1\Bigr)\,,\label{gdiff01}\\
d\log f &=&  d\log g + {{\kappa}\over{8\pi}} d\chi\nonumber\\
&=& {{\kappa r}\over{2 g}}\,\Bigl( {g\over{4\pi r}}\,d\chi -{1\over{4\pi r^2 g}}\,dm +{{m}\over{4\pi r^3}}\,\vartheta ^1\Bigr)\,,\nonumber\\
\label{fdiff01}
\end{eqnarray}
so that, using (\ref{Minksig00})-(\ref{Minksig11}), it follows
\begin{eqnarray}
e_0\rfloor d\log g &=& {{\kappa r}\over{2 g}}\,\Sigma _1{}^0\,,\label{gdiff02}\\
e_1\rfloor d\log g &=& {{\kappa r}\over{2 g}}\,\Bigl( \Sigma _0{}^0 + {{m}\over{4\pi r^3}}\Bigr)\,,\label{gdiff03}\\
e_1\rfloor d\log f &=& {{\kappa r}\over{2 g}}\,\Bigl( \Sigma _1{}^1 +{{m}\over{4\pi r^3}}\Bigr)\,.\label{fdiff02}
\end{eqnarray}
As shown by (\ref{Minksig00})-(\ref{fdiff02}), the energy-momentum matter source components in the rhs of (\ref{fieldeq1red}), calculated with the help of the spacetime functions $f$ and $g$ of the inner Schwarzschild metric (entering the gravitational energy-momentum form (\ref{Einstgravmom}) in the lhs of (\ref{fieldeq1red})), are fully described by the functions $m$ and $\chi$ and their derivatives.

\subsection{Conservation of the matter energy-momentum}

Being the gravitational energy-momentum (\ref{Einstgravmom}) identically conserved, the Einstein equation (\ref{fieldeq1red}) guarantees the automatic conservation of the material energy-momentum. Actually, according to (\ref{sigmamattconserv}) with zero torsion and no spin current
\begin{equation}
D\,\Sigma _\alpha = 0\,.\label{sigmamattconservred}
\end{equation}
We will show explicitly how it occurs by expressing (\ref{sigmamattconservred}) in the form
\begin{equation}
0=D\Sigma _\alpha =D\bigl(\Sigma _\alpha{}^\mu \eta _\mu\bigr) = D\Sigma _\alpha{}^\mu \wedge\eta _\mu = \bigl( e_\mu\rfloor D\Sigma _\alpha{}^\mu \bigr)\eta\,.\label{explenmomcons01}
\end{equation}
Let us first evaluate the radial equation, being careful to include the non-zero contributions of covariant derivatives of vanishing energy-momentum components as
\begin{eqnarray}
e_\mu\rfloor D\Sigma _1{}^\mu &=& e_0\rfloor D\Sigma _1{}^0 + e_1\rfloor D\Sigma _1{}^1 + e_2\rfloor D\Sigma _1{}^2 + e_3\rfloor D\Sigma _1{}^3\nonumber\\
&=&  e_0\rfloor\bigl[d\Sigma _1{}^0 +\Gamma _1{}^0\bigl(\Sigma _1{}^1 -\Sigma _0{}^0\bigr)\bigr]\nonumber\\
&& + e_1\rfloor \bigl( d\Sigma _1{}^1 + 2\,\Gamma _0{}^1\Sigma _1{}^0\bigr)\nonumber\\
&& + e_2\rfloor \bigl[\,\Gamma _1{}^2\bigl(\Sigma _1{}^1 -\Sigma _2{}^2\bigr)\bigr]\nonumber\\
&& + e_3\rfloor \bigl[\,\Gamma _1{}^3\bigl(\Sigma _1{}^1 -\Sigma _3{}^3\bigr)\bigr]\,.\label{explenmomcons02}
\end{eqnarray}
Replacing (\ref{Christoff01})-(\ref{Christoff05}) in (\ref{explenmomcons02}), the latter becomes
\begin{eqnarray}
e_\mu\rfloor D\Sigma _1{}^\mu &=& e_0\rfloor d\Sigma _1{}^0 + e_1\rfloor d\Sigma _1{}^1 -2\,(e_0\rfloor d\log{g})\Sigma _1{}^0\nonumber\\
&&+(e_1\rfloor d\log{f})\bigl(\Sigma _1{}^1 -\Sigma _0{}^0\bigr) + {{2g}\over{r}}\bigl(\Sigma _1{}^1 -\Sigma _2{}^2\bigr)\,,\nonumber\\
\label{Minksig22bis}
\end{eqnarray}
whose vanishing is identical with Eqn.(\ref{Minksig22}) for $\Sigma _2{}^2$. Proceeding in a similar way with the time equation, we find
\begin{eqnarray}
e_\mu\rfloor D\Sigma _0{}^\mu &=& e_0\rfloor d\Sigma _0{}^0 + e_1\rfloor d\Sigma _0{}^1 -2\,(e_1\rfloor d\log{f})\Sigma _1{}^0\nonumber\\
&&+(e_0\rfloor d\log{g})\bigl(\Sigma _1{}^1 -\Sigma _0{}^0\bigr) +{{2g}\over{r}}\Sigma _0{}^1\,.\nonumber\\
\label{explenmomcons04}
\end{eqnarray}
A simple calculation shows that it vanishes as a direct consequence of the values (\ref{Minksig00})--(\ref{Minksig11}) provided by the field equations. Finally, the angular equations are trivial
\begin{eqnarray}
e_\mu\rfloor D\Sigma _2{}^\mu &=& e_2\rfloor d\Sigma _2{}^2 =0\,,\label{explenmomcons05}\\
e_\mu\rfloor D\Sigma _3{}^\mu &=& e_3\rfloor d\Sigma _3{}^3 =0\,,\label{explenmomcons06}
\end{eqnarray}
since $\Sigma _2{}^2$ is a function of $t$ and $r$ only, with zero derivatives with respect to $\theta$ and $\varphi$, see (\ref{fr03}) and (\ref{fr04}). For completness, let us show the differential condition on the mass function $m$ which is hidden in the equations already considered. We achieve it by expressing the conservation of energy-momentum as
\begin{eqnarray}
D\Sigma _0 &=& d\Bigl( \Sigma _0{}^0\eta _0 + \Sigma _0{}^1\eta _1\Bigr) + d\log g\wedge\Sigma _1{}^1\eta _0\nonumber\\
&&- d\log f\wedge\Sigma _1{}^0\eta _1 =0\,,\label{consem00}\\
D\Sigma _1 &=& d\Bigl( \Sigma _1{}^0\eta _0 + \Sigma _1{}^1\eta _1\Bigr) - d\log g\wedge\Sigma _1{}^0\eta _0\nonumber\\
&&- d\log f\wedge\Sigma _0{}^0\eta _1 - {{2g}\over{r}}\,\Sigma _2{}^2\eta =0\,,\label{consem01}\\
D\Sigma _2 &=& d\Bigl( \Sigma _2{}^2\eta _2\Bigr) - {{\cos\theta}\over{r\sin\theta}}\,\Sigma _2{}^2\eta\nonumber\\
&=& d\Sigma _2{}^2\wedge\eta _2 =0\,,\label{consem02}\\
D\Sigma _3 &=& d\Bigl( \Sigma _2{}^2\eta _3\Bigr) = d\Sigma _2{}^2\wedge\eta _3 =0\,.\label{consem03}
\end{eqnarray}
Since (\ref{consem00}) is solved by (\ref{Minksig00})--(\ref{Minksig10}) and (\ref{consem02}) and (\ref{consem03}) follow from the functional dependence of $\Sigma _2{}^2$, let us look only for a more convenient form of (\ref{consem01}). From (\ref{Minksig00}) and (\ref{Minksig10}) we find out that
\begin{equation}
{1\over{4\pi r^2 g}}\,{}^* dm = \Sigma _1{}^0\eta _0 - \Sigma _0{}^0\eta _1 \,,\label{mdiff02}
\end{equation}
and consequently
\begin{equation}
{1\over{\bigl( 4\pi r^2 g\bigr)^2}}\,dm\wedge {}^* dm = -\Bigl[ \Bigl( \Sigma _1{}^0\Bigr)^2 - \Bigl( \Sigma _0{}^0\Bigr)^2\Bigr]\eta\,.\label{meandens06}
\end{equation}
The components of the energy-momentum 3-form (\ref{enmom01}) as derived from (\ref{Minksig00})-(\ref{Minksig33}) with (\ref{mdiff02}) read in compact form
\begin{eqnarray}
\Sigma _0 &=& -{1\over{4\pi r^2 g}}\,dm\wedge\eta _{01}\,,\label{sigma0}\\
\Sigma _1 &=& {1\over{4\pi r^2 g}}\,{}^*dm + \bigl( \Sigma _0{}^0 + \Sigma _1{}^1\bigr)\eta _1\,,\label{sigma1bis}\\
\Sigma _2 &=& \Sigma _2{}^2\,\eta _2\,,\label{sigma2}\\
\Sigma _3 &=& \Sigma _2{}^2\,\eta _3\,,\label{sigma3}
\end{eqnarray}
so that Eq.(\ref{consem01}) with (\ref{mdiff02}), (\ref{sigma1bis}) and (\ref{gdiff02})-(\ref{fdiff02}) takes the form
\begin{eqnarray}
0 &=& d\Bigl[ {1\over{4\pi r^2 g}}\,{}^* dm + \bigl( \Sigma _0{}^0 + \Sigma _1{}^1\bigr)\eta _1\Bigr]\nonumber\\
&& + {{\kappa r}\over{2g}}{1\over{\bigl( 4\pi r^2 g\bigr)}}\,dm\wedge\Bigl( \Sigma _0{}^0 + \Sigma _1{}^1 + {{m}\over{4\pi r^3}}\Bigr)\eta _1\nonumber\\
&& - {{\kappa r}\over{2g}}\Bigl[ \Bigl( \Sigma _1{}^0\Bigr)^2 - \Bigl( \Sigma _0{}^0\Bigr)^2\Bigr]\eta - {{2g}\over{r}} \Sigma _2{}^2\,\eta\,.\label{meandens03}
\end{eqnarray}
Operating and taking into account (\ref{meandens06}), we finally get
\begin{eqnarray}
0 &=& {1\over{4\pi r^2 g}}\,d{}^* dm + {{\kappa r}\over{g}}{1\over{\bigl( 4\pi r^2 g\bigr)^2}}\,dm\wedge {}^* dm\nonumber\\
&& +{{\kappa r}\over{2g}}{1\over{\bigl( 4\pi r^2 g\bigr)}}\,dm\wedge\Bigl( \Sigma _0{}^0 + \Sigma _1{}^1\Bigr)\eta _1\nonumber\\
&& + d\Bigl[\Bigl( \Sigma _0{}^0 + \Sigma _1{}^1\Bigr)\eta _1\Bigr] + {{2g}\over{r}}\bigl( \Sigma _0{}^0 - \Sigma _2{}^2\bigl)\,\eta\,.\nonumber\\
\label{meandens03bis}
\end{eqnarray}
Our next task will be to penetrate the inner structure of the previous general results with the aim of obtaining information concerning the different pieces of Eckart's decomposition of the energy-momentum tensor.

\subsection{Identifying the energy-momentum constituents}

In view of the symmetry condition (\ref{symmsigma01}) following from vanishing spin current, the energy-momentum tensor (\ref{enmom02}) simplifies to the symmetric expression
\begin{equation}
\Sigma _\alpha{}^\beta= \rho\,u_\alpha u^\beta + p\,h_\alpha{}^\beta -u_\alpha q^\beta -q_\alpha u^\beta + S_\alpha{}^\beta\,,\label{symmenmom02}
\end{equation}
(with $p_\alpha = q_\alpha$ and $\Pi _\alpha{}^\beta = S_\alpha{}^\beta$), and the energy and momentum conservation equations (\ref{mattender03}) and (\ref{mattmomder04}) reduce respectively to
\begin{equation}
{\it l}_u (\rho\overline{\eta}) -\underline{d}\,(q^\beta\overline{\eta}_\beta ) + p\,{\it l}_u \overline{\eta} - q_\alpha\,{\cal \L\/}_u u^\alpha\,\overline{\eta}+ \underline{D}u^\alpha\wedge S_\alpha{}^\beta\,\overline{\eta}_\beta =0\,,\label{redmattender04}
\end{equation}
and
\begin{equation}
h_\alpha{}^\beta \Bigl[ {\cal \L\/}_u\bigl( q_\beta\,\overline{\eta}\bigr) - \underline{D}\bigl( S_\beta{}^\gamma\,\overline{\eta}_\gamma\bigr)\Bigr] - \underline{d}p\wedge\overline{\eta}_\alpha = -\underline{D} u_\alpha\wedge q^\beta\overline{\eta}_\beta\,.\label{redmattender05}
\end{equation}
Notice that, in both equations, the contribution of the viscosity stress tensor $S_\alpha{}^\beta$ can be split into shear and bulk viscosities according to (\ref{tracefreeS01}), in such a way that the pressure $p$ transforms into the effective pressure
\begin{equation}
p_{\rm eff}:= p +{1\over 3} S_\mu{}^\mu\,,\label{effpressure01}
\end{equation}
showing the role of $S_\mu{}^\mu$ as compression. Although the identification of phenomenological pressure with this quantity remains an open possibility, here we will consider the bulk viscosity as a separate variable. The interpretation of the dynamical content of (\ref{redmattender04}) and (\ref{redmattender05}) requires to identify the basic pieces (\ref{rho01})--(\ref{pressandS01}) of the spherical source (\ref{symmenmom02}) in terms of the components (\ref{Minksig00})--(\ref{Minksig33}) derived from the Einstein equations. The obvious guide to look for the constitutive matter variables is provided by their defining equations (\ref{rho01})-(\ref{Strace01}) themselves, which, in the present case, read
\begin{eqnarray}
&&\rho = -2\,\Sigma _1{}^0 u_1 u^0 -(u^0)^2\,\Sigma _0{}^0 + u_1^2\,\Sigma _1{}^1 + \bigl( u_2^2 +u_3^2\bigr)\Sigma _2{}^2\,,\nonumber\\
\label{rho01bis}\\
&&q_\alpha = h_\alpha{}^0\Bigl[\bigl( \Sigma _0{}^0 -\Sigma _2{}^2\bigr) u_0 + \Sigma _0{}^1 u_1\Bigr]\nonumber\\
&&\hskip0.8cm + h_\alpha{}^1\Bigl[\bigl( \Sigma _1{}^1 -\Sigma _2{}^2\bigr) u_1 + \Sigma _1{}^0 u_0\Bigr]\,,\label{flux02}\\
&&\Sigma _\mu{}^\mu = -\rho +3p +S_\mu{}^\mu\,,\label{pressandS02}\\
&&{\slash\!\!\!\!S}_\alpha{}^\beta = h_\alpha{}^\beta \Bigl[\Sigma _2{}^2 -{1\over 3}\,\bigl( \rho + \Sigma _\mu{}^\mu \bigr)\Bigr]\nonumber\\
&&\hskip1.4cm + h_\alpha{}^0 \Bigl[\bigl( \Sigma _0{}^0 -\Sigma _2{}^2\bigr) h_0{}^\beta + \Sigma _0{}^1 h_1{}^\beta\Bigr]\nonumber\\
&&\hskip1.4cm + h_\alpha{}^1 \Bigl[\bigl( \Sigma _1{}^1 -\Sigma _2{}^2\bigr) h_1{}^\beta + \Sigma _1{}^0 h_0{}^\beta\Bigr]\,.\label{tracelessS02}
\end{eqnarray}
We shall proceed in two steps. First we relate the physical quantities in the rhs of (\ref{symmenmom02}) to the inner Schwarzschild functions $m$ and $\chi$ (whose possible dependence on other variables such as mass density or temperature, etc. remains to be determined) and later we will describe, with the help of these constitutive element of (\ref{symmenmom02}), a particular spherical matter source --a star, say-- composed of a Van der Waals fluid and radiation characterized as a photon gas. Let us begin by considering Eq.(\ref{rho01bis}), trivially reformulated as
\begin{eqnarray}
\rho +\Sigma _0{}^0 &=& -2\,\Sigma _1{}^0 u_1 u^0 +\bigl(\Sigma _1{}^1 -\Sigma _2{}^2\bigr) u_1^2\nonumber\\
&& -\bigl(\Sigma _2{}^2 -\Sigma _0{}^0\bigr) h_0{}^0\,.\label{rho01modbis}
\end{eqnarray}
Taking into account (\ref{mdiff02}), whose decomposition into longitudinal and transversal parts yields
\begin{eqnarray}
{1\over{4\pi r^2 g}}\,{}^{\#}\underline{d}m &=& \Sigma _1{}^0\overline{\eta}_0 - \Sigma _0{}^0\overline{\eta}_1\,,\label{mdertrans}\\
{1\over{4\pi r^2 g}}\,{\it l}_u m &=& - \bigl( \Sigma _1{}^0 u^0 + \Sigma _0{}^0 u_1\bigr)\,,\label{mderlong}
\end{eqnarray}
by replacing (\ref{mderlong}) in (\ref{rho01modbis}) we get the information contained in the latter expressed as
\begin{eqnarray}
{{2 u_1}\over{4\pi r^2 g}}\,{\it l}_u m &=& \bigl( \rho +\Sigma _0{}^0\bigr) -\bigl( 2\,\Sigma _0{}^0 +\Sigma _1{}^1 -\Sigma _2{}^2\bigr) u_1^2\nonumber\\
&& +\bigl(\Sigma _2{}^2 -\Sigma _0{}^0\bigr) h_0{}^0\,.\label{rho03}
\end{eqnarray}
On the other hand, from (\ref{flux02}) we build the energy flux 2-form
\begin{eqnarray}
q^\alpha\overline{\eta}_\alpha &=& \Bigl[\bigl( \Sigma _0{}^0 -\Sigma _2{}^2\bigr) u^0 + \Sigma _1{}^0 u_1\Bigr]\overline{\eta}_0\nonumber\\
&& + \Bigl[\bigl( \Sigma _1{}^1 -\Sigma _2{}^2\bigr) u_1 -\Sigma _1{}^0 u^0\Bigr]\overline{\eta}_1\,,\label{enflow01}
\end{eqnarray}
and rewrite it as
\begin{eqnarray}
q^\alpha\overline{\eta}_\alpha &=& u_1\bigl( \Sigma _1{}^0\overline{\eta}_0 - \Sigma _0{}^0\overline{\eta}_1\bigr) -\bigl( \Sigma _1{}^0 u^0 +\Sigma _0{}^0 u_1\bigr)\overline{\eta}_1\nonumber\\
&& + \bigl( \Sigma _0{}^0 +\Sigma _1{}^1\bigr) u_1\overline{\eta}_1 + \bigl( \Sigma _2{}^2 -\Sigma _0{}^0\bigr)\bigl( u_2\overline{\eta}_2 + u_3\overline{\eta}_3\bigr)\,,\nonumber\\
\label{enflow02}
\end{eqnarray}
so that, by replacing in it (\ref{mdertrans}) and (\ref{mderlong}), it becomes
\begin{eqnarray}
q^\alpha\overline{\eta}_\alpha &=& {{u_1}\over{4\pi r^2 g}}\,{}^{\#}\underline{d}m +{1\over{4\pi r^2 g}}\,{\it l}_u m\,\overline{\eta}_1 + \bigl( \Sigma _0{}^0 +\Sigma _1{}^1\bigr) u_1\overline{\eta}_1\nonumber\\
&& + \bigl( \Sigma _2{}^2 -\Sigma _0{}^0\bigr)\bigl( u_2\overline{\eta}_2 + u_3\overline{\eta}_3\bigr)\,.\label{enflow03}
\end{eqnarray}
Eliminating the Lie derivative term from (\ref{enflow03}) by means of (\ref{rho03}), we get
\begin{eqnarray}
q^\alpha\overline{\eta}_\alpha &=& {{u_1}\over{4\pi r^2 g}}\,{}^{\#}\underline{d}m + \bigl(\Sigma _\mu{}^\mu - 3\Sigma _2{}^2\bigr) u_1\overline{\eta}_1\nonumber\\
&& -\bigl(\Sigma _2{}^2 -\Sigma _0{}^0\bigr) u^0\overline{\eta}_0\nonumber\\
&& + \Bigl[\bigl( \rho +\Sigma _0{}^0\bigr) -\bigl( \Sigma _1{}^1 -\Sigma _2{}^2\bigr)u_1^2\nonumber\\
&&\hskip2.1cm +\bigl(\Sigma _2{}^2 -\Sigma _0{}^0\bigr) h_0{}^0\Bigr] {1\over{2 u_1}}\overline{\eta}_1\,.\nonumber\\
\label{enflow04}
\end{eqnarray}
The fact that the energy flux appears as a gradient term plus additional contributions resembles formally the Fourier law of thermal conduction and Fick's first diffusion law. Notwithstanding, a different formulation of the same quantity $q_\alpha$ is possible, as we will see later.

Let us now invoke Eq.(\ref{pressandS02}). Consistently with it, for the diagonal terms of the energy-momentum tensor (\ref{symmenmom02}) we make the {\it ansatz}
\begin{eqnarray}
\Sigma _0{}^0 &=& -\rho\,,\label{assumpt00simplif}\\
\Sigma _1{}^1 &=& p + S_\mu{}^\mu\,,\label{assumpt11simplif}\\
\Sigma _2{}^2 &=& p\,,\label{assumpt22simplif}\\
\Sigma _3{}^3 &=& p\,,\label{assumpt33simplif}
\end{eqnarray}
and from (\ref{rho01modbis}) with (\ref{assumpt00simplif})-(\ref{assumpt33simplif}) we find the remaining non vanishing component to be
\begin{equation}
\Sigma _1{}^0 = -{1\over{2u_1 u^0}}\Bigl[ \bigl(\rho + p\bigr) h_0{}^0 - S_\mu{}^\mu u_1^2\Bigr]\,.\label{explictrace05simplif}
\end{equation}
With these choices, (\ref{rho03}) reduces to
\begin{equation}
{{2 u_1}\over{4\pi r^2 g}}\,{\it l}_u m = \bigl( 2\rho - S_\mu{}^\mu\bigr) u_1^2 +\bigl(\rho + p\bigr) h_0{}^0\,,\label{rho03explic}
\end{equation}
and Eq.(\ref{enflow04}) for the energy flux becomes
\begin{eqnarray}
q^\alpha\overline{\eta}_\alpha &=& {{u_1}\over{4\pi r^2 g}}\,{}^{\#}\underline{d}m + \bigl(\Sigma _\mu{}^\mu - 3p\bigr) u_1\overline{\eta}_1 -\bigl(\rho + p\bigr) u^0\overline{\eta}_0\nonumber\\
&& + \Bigl[\bigl(\rho + p\bigr) h_0{}^0 -S_\mu{}^\mu u_1^2\Bigr] {1\over{2 u_1}}\overline{\eta}_1\,.\label{enflow05}
\end{eqnarray}
Finally, the shear viscous stress tensor as calculated from (\ref{tracelessS02}) with (\ref{assumpt00simplif})-(\ref{explictrace05simplif}) reads
\begin{eqnarray}
{\slash\!\!\!\!S}_\alpha{}^\beta &=& S_\mu{}^\mu\,\Bigl\{ -{1\over 3} h_\alpha{}^\beta + h_\alpha{}^1 h_1{}^\beta + {{u_1}\over{2 u^0}}\Bigl( h_\alpha{}^1 h_0{}^\beta - h_\alpha{}^0 h_1{}^\beta\Bigr)\Bigr\}\nonumber\\
&& -\bigl(\rho +p\bigr)\Bigl\{ h_\alpha{}^0 h_0{}^\beta + {{h_0{}^0}\over{2 u_1 u^0}}\,\Bigl( h_\alpha{}^1 h_0{}^\beta - h_\alpha{}^0 h_1{}^\beta\Bigr)\Bigr\}\,.\nonumber\\
\label{tracelessS02simplif}
\end{eqnarray}
Eqs.(\ref{assumpt00simplif})-(\ref{assumpt33simplif}) have also consequences for the remaining constitutive elements of energy-momentum. From (\ref{Minksig00}) and (\ref{assumpt00simplif}) we read out
\begin{equation}
\rho = {1\over{4\pi r^2 g}}\,(e_1\rfloor dm)\,,\label{Minksig00simplif}
\end{equation}
in such a way that the mass definition (\ref{gfunct02}) takes the form
\begin{equation}
m:= \int _0^r \rho\,4\pi r^2 dr\,.\label{exlicitmass01}
\end{equation}
Analogously, Eq.(\ref{Minksig11}) with (\ref{assumpt00simplif}) and (\ref{assumpt11simplif}) becomes
\begin{equation}
\rho + p + S_\mu{}^\mu = {g\over{4\pi r}}\,(e_1\rfloor d\chi) \,,\label{pressure01}
\end{equation}
involving density, pressure and bulk viscosity, so that (\ref{ffunct03}) transforms into
\begin{equation}
\chi := \int _0^r\Bigl( \rho + p + S_\mu{}^\mu\Bigr){2\over{g^2}} 2\pi r dr\,.\label{ffunct03bis}
\end{equation}
Finally, (\ref{mdiff02}) with (\ref{assumpt00simplif}) and (\ref{flux04}) below, reads
\begin{equation}
{1\over{4\pi r^2 g}}\,{}^* dm = {{q^0}\over{u_1}}\eta _0 +\rho\eta _1\,,\label{mdiff03}
\end{equation}
which, replaced in (\ref{meandens03}) together with (\ref{assumpt00simplif})-(\ref{assumpt33simplif}), yields
\begin{eqnarray}
0 &=& d \Bigl[ {{q^0}\over{u_1}}\eta _0 + \bigl( p + S_\mu{}^\mu\bigr)\eta _1\Bigr] - {{2g}\over{r}} p\,\eta\nonumber\\
&& + {{\kappa r}\over{2g}} \Bigl[ -\Bigl( {{q^0}\over{u_1}}\Bigr)^2 + \rho\,\bigl( p + S_\mu{}^\mu + {{m}\over{4\pi r^3}}\bigr)\Bigr]\eta\,.\nonumber\\
\label{qzeroder}
\end{eqnarray}
This completes the information provided by gauge gravity alone. However, a further development is allowed by taking into account the thermodynamical approach presented in Sect.VII, with the auxiliary variables introduced there.

\subsection{Link to thermodynamical matter models}

As pointed out in Sect II, the sources of gravity must include all material contributions due to both, matter proper and radiation. We will consider a gaseous sphere whose material and radiative aspects are described by a Van der Waals fluid and a photon gas respectively, providing the basic constituents necessary to develop a Poincar\'e gauge theoretical model of the structure and evolution of a non-rotating star. The total energy-momentum is the sum of the fluid and radiation pieces
\begin{equation}
\Sigma _\alpha = \Sigma ^{\rm VdW}_\alpha +\Sigma ^{\rm ph}_\alpha\,,\label{starem01}
\end{equation}
which can be studied separately by assuming that possible no gravitational forces $D\Sigma ^{\rm VdW}_\alpha = f_\alpha$ cancel out from the global condition $D\Sigma _\alpha =0$ to be considered at last, see Sect.II. Several contributions to (\ref{starem01}) are known from the solution of the Gibbs equations (\ref{ansatz03}) and (\ref{ansatz04}). Among the constitutive elements (\ref{symmenmom02}) of (\ref{starem01}), the total energy density $\rho$ is given by (\ref{rhodecomp01}) with the internal energy $\mathsf{u}$ as the sum of (\ref{VdWgasenergydens02}) and (\ref{photgasenergydens01}), and the total pressure $p$ is found by adding up the partial pressures (\ref{VdWgaspress02}) and (\ref{photgaspress01}). Notice that (\ref{exlicitmass01}) with (\ref{rhodecomp01}) reads
\begin{equation}
m:= \int _0^r \bigl(\rho _{\rm m}+\mathsf{u}\bigr)\,4\pi r^2 dr\,,\label{exlicitmass01bis}
\end{equation}
showing how the internal energy contributes to the total gravitational mass. However, the known derivatives (\ref{mdertrans}) and (\ref{mderlong}) of (\ref{exlicitmass01bis}), expressed in terms of the quantity $\Sigma _1{}^0$ which is not determined by the Gibbs equation (\ref{uLiederiv02}), are useless to calculate the energy flow (\ref{enflow05}) for the particular model (Van der Waals plus thermal photons) we are interested in. In order to overcome this difficulty, I propose to introduce a speculative conjecture allowing to find the form of (\ref{enflow05}) corresponding to any kind of phenomenological matter. Let us make use of the relation
\begin{equation}
{{r}\over{g}}\,{}^* d\Bigl( {{m}\over{4\pi r^3}}\Bigr) \equiv {1\over{4\pi r^2 g}}\,{}^* dm - {{3 m}\over{4\pi r^3}}\eta _1\,,\label{meandens04}
\end{equation}
involving the mean mass density ${{3 m}\over{4\pi r^3}}$. Decomposing (\ref{meandens04}) into
\begin{eqnarray}
{1\over{4\pi r^2 g}}\,{}^{\#}\underline{d}m &=& {{r}\over{g}}\,{}^{\#}\underline{d}\Bigl( {{m}\over{4\pi r^3}}\Bigr) +{{3 m}\over{4\pi r^3}}\overline{\eta}_1\,,\label{identmdertrans}\\
{1\over{4\pi r^2 g}}\,{\it l}_u m &=& {{r}\over{g}}\,{\it l}_u \Bigl( {{m}\over{4\pi r^3}}\Bigr) +{{3 m}\over{4\pi r^3}} u_1\,,\label{identmderlong}
\end{eqnarray}
and replacing (\ref{identmdertrans}) in (\ref{enflow05}), one gets
\begin{eqnarray}
q^\alpha\overline{\eta}_\alpha &=& {{u_1 r}\over{3g}}\,{}^{\#}\underline{d}\Bigl( {{3m}\over{4\pi r^3}}\Bigr) + \Bigl( {{3 m}\over{4\pi r^3}} +\Sigma _\mu{}^\mu - 3p\Bigr) u_1\overline{\eta}_1\nonumber\\
&& -\bigl(\rho + p\bigr) u^0\overline{\eta}_0 +\Bigl[\bigl(\rho + p\bigr) h_0{}^0 -S_\mu{}^\mu u_1^2\Bigr] {1\over{2 u_1}}\overline{\eta}_1\,.\nonumber\\
\label{enflow06}
\end{eqnarray}
My non deductively justified assumption is suggested by analogy with (\ref{Direnmomtrace}) and (\ref{Bostrace}). It consists in choosing the trace of the energy-momentum tensor to be
\begin{equation}
\Sigma _\mu{}^\mu = -{{3 m}\over{4\pi r^3}} + k p\,,\label{mattemtrace01}
\end{equation}
leaving the numerical constant $k$ unfixed so that it can take different values depending on the kind of matter considered. In support of (\ref{mattemtrace01}), notice the natural way in which the quantities involved in it are related. See for instance the second term in the rhs of (\ref{enflow06}). Furthermore, from (\ref{Minksig00}) and (\ref{Minksig11}) with (\ref{gdiff03}) and (\ref{fdiff02}) we find
\begin{eqnarray}
\Sigma _\mu{}^\mu &=& -{2\over{4\pi r^2 g}}\,(e_1\rfloor dm) + {g\over{4\pi r}}\,(e_1\rfloor d\chi) + 2p\nonumber\\
&=& {{2g}\over{\kappa r}}\,e_1\rfloor d\log \bigl( fg\bigr) -{{2m}\over{4\pi r^3}} + 2p\,,\label{mattemtraceaux01}
\end{eqnarray}
implying that (\ref{mattemtrace01}) is equivalent to the condition
\begin{equation}
{{6g}\over{\kappa r}}\,e_1\rfloor d\log \bigl( fg\bigr) = \Sigma _\mu{}^\mu + 2\bigl( k-3\bigr) p\,.\label{mattemtraceaux02}
\end{equation}
In view of (\ref{Strace01}), the mean density in (\ref{mattemtrace01}) reads
\begin{equation}
{{3 m}\over{4\pi r^3}} = \rho + \bigl( k -3\bigr)p -S_\mu{}^\mu\,,\label{meandens10}
\end{equation}
where at least $\rho$ and $p$ are known from thermodynamics for different phenomenological models, as mentioned above. Making use of the hypothesis (\ref{mattemtrace01}), Eq.(\ref{enflow06}) takes the form
\begin{eqnarray}
q^\alpha\overline{\eta}_\alpha &=& {{u_1 r}\over{3g}}\,{}^{\#}\underline{d}\Bigl( {{3m}\over{4\pi r^3}}\Bigr) -\bigl(\rho + p\bigr) u^0\overline{\eta}_0\nonumber\\
&& +\Bigl[\bigl(\rho + p\bigr) h_0{}^0 -S_\mu{}^\mu u_1^2 + 2\bigl( k-3\bigr) p u_1^2\Bigr] {1\over{2 u_1}}\overline{\eta}_1\,,\nonumber\\
\label{enflow08}
\end{eqnarray}
involving the gradient of (\ref{meandens10}). Once the energy flux has become calculable, it still remains to identify its different pieces (\ref{energfluxdecomp01}) in order to make it possible to perform in practice the reduction of the energy conservation equation (\ref{redmattender04}) to the first law of thermodynamics (\ref{firstlawtherm01}). Let us rewrite (\ref{energfluxdecomp01}) for a single particle species as
\begin{equation}
q^\alpha\overline{\eta}_\alpha = \mathfrak{q}_{_H} -\mathbb{J}_{\rm m} -\mu \mathcal{J}\,,\label{energfluxdecomp02}
\end{equation}
and replace (\ref{rhodecomp01}) in (\ref{meandens10}) to get
\begin{equation}
{{3 m}\over{4\pi r^3}} = \rho _{\rm m} + \mathsf{u} +\bigl( k-3\bigr)p -S_\mu{}^\mu\,.\label{meandens11}
\end{equation}
Our attempt is to separate variables into the distinguished pieces of the rhs of (\ref{energfluxdecomp02}). Assuming the bulk viscosity not to depend on the mass density $\rho _{\rm m}$, we identify the mass current as
\begin{equation}
-\mathbb{J}_{\rm m} := {u_1 r\over{3g}}\,{}^{\#}\underline{d}\rho _{\rm m} +\rho _{\rm m}\Bigl( -u^0\overline{\eta}_0 + {{h_0{}^0}\over{2u_1}}\overline{\eta}_1\Bigr)\,,\label{enflux03}
\end{equation}
while the sum of heat and particle number flows reads
\begin{eqnarray}
\mathfrak{q}_{_H} -\mu \mathcal{J} &:=& {u_1 r\over{3g}}\,{}^{\#}\underline{d}\bigl[\,\mathsf{u} +\bigl( k-3\bigr)p -S_\mu{}^\mu\,\bigr] -{1\over 2}\,S_\mu{}^\mu u_1\overline{\eta}_1\nonumber\\
&& + \bigl(\mathsf{u} +p\bigr)\Bigl[ -u^0\overline{\eta}_0 + {{h_0{}^0}\over{2u_1}}\overline{\eta}_1\Bigr] + \bigl( k -3\bigr)p\,u_1\overline{\eta}_1\,.\nonumber\\
\label{enflux04}
\end{eqnarray}
Let us try to identify the two pieces separately in the particular case of a Van der Waals fluid.

\subsection{Van der Waals energy flux pieces}

Van der Waals fluids posses internal energy (\ref{VdWgasenergydens02}) and pressure (\ref{VdWgaspress02}), so that Eq.(\ref{meandens11}), with the constant $k$ denoted as $k_1$ in order to distinguish its value from that of radiation to be studied below, reads
\begin{equation}
{{3 m}\over{4\pi r^3}} = \rho _{\rm m} -\bigl( k_1 -2\bigr) a n^2 + \beta T - S_\mu{}^\mu \,,\label{meandens12}
\end{equation}
where the function $\beta$ of particle number density $n$ (assumed to be independent of $T$) is defined as
\begin{equation}
\beta := \Bigl[ c_{\rm n} + {{\bigl( k_1 -3\bigr)}\over{\bigl(1 -b n\bigr)}}\Bigr] k_{_B} n\,.\label{beta01}
\end{equation}
We also rewrite the sum of (\ref{VdWgasenergydens02}) and (\ref{VdWgaspress02}) as
\begin{equation}
\mathsf{u} + p = \gamma T - 2a n^2\,,\label{VdW01}
\end{equation}
in terms of
\begin{equation}
\gamma := \Bigl[ c_{\rm n} + {1\over{\bigl(1 -b n\bigr)}}\Bigr] k_{_B} n\,.\label{gamma01}
\end{equation}
On the other hand, using (\ref{VdW01}) we find
\begin{equation}
\,{}^{\#}\underline{d}\bigl( a n^2\bigr) = 2a n^2\,{}^{\#}\underline{d}\log{n} = \bigl[\,\gamma T -\bigl( \mathsf{u} + p\bigr)\bigr]\,{}^{\#}\underline{d}\log{n}\,,\label{VdW02}
\end{equation}
and we express (\ref{VdWgaspress02}) as
\begin{equation}
p = {{k_{_B} n T}\over{\bigl(1 -b n\bigr)}}\, -{1\over 2}\bigl[\,\gamma T -\bigl( \mathsf{u} + p\bigr)\bigr]\,.\label{VdW03}
\end{equation}
Replacing (\ref{meandens12}), (\ref{VdW02}) and (\ref{VdW03}) in (\ref{enflux04}) (where (\ref{enflux03}) was already separated), one gets
\begin{eqnarray}
\mathfrak{q}_{_H} -\mu \mathcal{J} &:=& {u_1 r\over{3g}}\,{}^{\#}\underline{d}\bigl( \beta T -S_\mu{}^\mu\,\bigr) + {1\over 2}\bigl( \beta T -S_\mu{}^\mu\,\bigr)\,u_1\overline{\eta}_1\nonumber\\
&& - \bigl( k_1 -2\bigr)\Bigl[ \gamma T\,{u_1 r\over{3g}}\,{}^{\#}\underline{d}\log{n} + {1\over 2}\,c_{\rm n}k_{_B} n T\,u_1\overline{\eta}_1\Bigr]\nonumber\\
&& + \bigl(\mathsf{u} +p\bigr)\Bigl[ \bigl( k_1 -2\bigr){u_1 r\over{3g}}\,{}^{\#}\underline{d}\log{n}\nonumber\\
&&\hskip1.5cm -u^0\overline{\eta}_0 + {{h_0{}^0}\over{2u_1}}\overline{\eta}_1 + {1\over 2}\bigl( k_1 -3\bigr)\,u_1\overline{\eta}_1\Bigr]\,.\nonumber\\
\label{VdW04}
\end{eqnarray}
Taking into account (\ref{ansatz04}), the particle number flux is easily identified as
\begin{eqnarray}
-\mu\mathcal{J} &:=& \mu n\Bigl[ \bigl( k_1 -2\bigr){u_1 r\over{3g}}\,{}^{\#}\underline{d}\log{n}\nonumber\\
&&\hskip0.8cm -u^0\overline{\eta}_0 + {{h_0{}^0}\over{2u_1}}\overline{\eta}_1 + {1\over 2}\bigl( k_1 -3\bigr)\,u_1\overline{\eta}_1\Bigr]\,,\nonumber\\
\label{enflux14}
\end{eqnarray}
and the heat flow thus reads
\begin{eqnarray}
\mathfrak{q}_{_H}  &:=& {u_1 r\over{3g}}\,{}^{\#}\underline{d}\bigl( \beta T -S_\mu{}^\mu\,\bigr) + {1\over 2}\bigl( \beta T -S_\mu{}^\mu\,\bigr)\,u_1\overline{\eta}_1\nonumber\\
&& - \bigl( k_1 -2\bigr)\Bigl[ \gamma T\,{u_1 r\over{3g}}\,{}^{\#}\underline{d}\log{n} + {1\over 2}\,c_{\rm n}k_{_B} n T\,u_1\overline{\eta}_1\Bigr]\nonumber\\
&& - T\sigma {1\over n}\mathcal{J}\,.\label{enflux13}
\end{eqnarray}
Let us compare these results with the alternative formulation of the energy flow calculated from (\ref{flux03bis}) (or from (\ref{momdens02}), since $p_\alpha =q_\alpha$). In the present case, this condition takes the form
\begin{eqnarray}
q_0 &=& u_0 \bigl(\rho +\Sigma _0{}^0  \bigr) + \Sigma _0{}^1 u_1\,,\label{flux04}\\
q_1 &=& u_1 \bigl(\rho +\Sigma _1{}^1 \bigr) + \Sigma _1{}^0 u_0\,,\label{flux05}\\
q_2 &=& u_2 \bigl(\rho +\Sigma _2{}^2 \bigr)\,,\label{flux06}\\
q_3 &=& u_3 \bigl(\rho +\Sigma _2{}^2 \bigr)\,,\label{flux07}
\end{eqnarray}
which, with (\ref{assumpt00simplif})-(\ref{explictrace05simplif}), reduces to
\begin{eqnarray}
q_0 &=& {{1}\over{2u^0}}\Bigl[ \bigl(\rho + p\bigr) h_0{}^0 - S_\mu{}^\mu u_1^2\Bigr]\,,\label{flux08}\\
q_1 &=& u_1\bigl(\rho + p\bigr) +{{1}\over{2u_1}}\Bigl[ \bigl(\rho + p\bigr) h_0{}^0 + S_\mu{}^\mu u_1^2\Bigr]\,,\label{flux09}\\
q_2 &=& u_2 \bigl(\rho + p\bigr)\,,\label{flux10}\\
q_3 &=& u_3 \bigl(\rho + p\bigr)\,.\label{flux11}
\end{eqnarray}
The decomposition into separate pieces (\ref{energfluxdecomp02}) is immediate by using (\ref{rhodecomp01}) and (\ref{ansatz04}) to show that
\begin{eqnarray}
\rho + p &=& \rho _{\rm m} + \mathsf{u} + p\,,\label{rhoplusp01}\\
&=& \rho _{\rm m} + T\sigma +\mu n\,,\label{rhopluspdecomp01}
\end{eqnarray}
allowing to identify the mass flow (or mechanical momentum) components
\begin{eqnarray}
-\mathsf{J}^{\rm m}_0 &=& \rho _{\rm m}\,{{h_0{}^0}\over{2 u^0}}\,,\label{masscurrcomp0}\\
-\mathsf{J}^{\rm m}_1 &=& \rho _{\rm m}\Bigl( u_1 + {{h_0{}^0}\over{2 u_1}}\Bigr)\,,\label{masscurrcomp1}\\
-\mathsf{J}^{\rm m}_2 &=& \rho _{\rm m} u_2\,,\label{masscurrcomp2}\\
-\mathsf{J}^{\rm m}_3 &=& \rho _{\rm m} u_3\,,\label{masscurrcomp3}
\end{eqnarray}
as much as the components of the particle number flux
\begin{eqnarray}
-J _0 &=& n\,{{h_0{}^0}\over{2 u^0}}\,,\label{partcurrcomp0}\\
-J _1 &=& n\Bigl( u_1 + {{h_0{}^0}\over{2 u_1}}\Bigr)\,,\label{partcurrcomp1}\\
-J _2 &=& n\,u_2\,,\label{partcurrcomp2}\\
-J _3 &=& n\,u_3\,,\label{partcurrcomp3}
\end{eqnarray}
and those of the heat flux
\begin{eqnarray}
q^{_H}_0 &=& T\sigma\,{{h_0{}^0}\over{2 u^0}} - S_\mu{}^\mu\,{{u_1^2}\over{2 u^0}}\,,\label{heatcurrcomp0}\\
q^{_H}_1 &=& T\sigma\Bigl( u_1 + {{h_0{}^0}\over{2 u_1}}\Bigr) +{1\over 2} S_\mu{}^\mu u_1\,,\label{heatcurrcomp1}\\
q^{_H}_2 &=& T\sigma\,u_2\,,\label{heatcurrcomp2}\\
q^{_H} _3 &=& T\sigma\,u_3\,.\label{heatcurrcomp3}
\end{eqnarray}
The energy flow 2-form built from the components (\ref{flux08})-(\ref{flux11}) reads
\begin{eqnarray}
q^\alpha\overline{\eta}_\alpha &=& -\bigl(\rho + p\bigr)\Bigl[\,u^0\overline{\eta}_0 + h_0{}^0 \Bigl( {1\over{2 u^0}}\overline{\eta}_0 -{1\over{2 u_1}}\overline{\eta}_1\Bigr)\Bigr]\nonumber\\
&& + S_\mu{}^\mu u_1^2\,\Bigl( {1\over{2 u^0}}\overline{\eta}_0 +{1\over{2 u_1}}\overline{\eta}_1\Bigr)\,.\label{nogradenflow01}
\end{eqnarray}
The link between this simple expression and the previously obtained ones involving gradients is established by means of (\ref{mdertrans}), which in view of (\ref{assumpt00simplif}) and (\ref{flux04}) relates to the first flux component as
\begin{equation}
{{u_1}\over{4\pi r^2 g}}\,{}^{\#}\underline{d}m = q^0\overline{\eta}_0 +\rho\,u_1\overline{\eta}_1\,.\label{gradform01}
\end{equation}
Returning to the notation (\ref{mdertrans}) and replacing in it the explicit form (\ref{explictrace05simplif}) of $\Sigma _1{}^0$ as much as (\ref{assumpt00simplif}), and taking into account (\ref{identmdertrans}) and (\ref{meandens10}), we get (\ref{gradform01}) transformed into
\begin{eqnarray}
{{u_1 r}\over{3g}}\,{}^{\#}\underline{d}\Bigl( {{3m}\over{4\pi r^3}}\Bigr) &=& -\Bigl[ \bigl(\rho + p\bigr) h_0{}^0 - S_\mu{}^\mu u_1^2\Bigr]{1\over{2 u^0}}\overline{\eta}_0\nonumber\\
&& + \Bigl[ S_\mu{}^\mu - \bigl( k_1 -3\bigr)p\Bigr] u_1\overline{\eta}_1\,.\label{gradform02}
\end{eqnarray}
Let us use (\ref{gradform02}) with (\ref{meandens12}) and (\ref{VdW01}), separating the different variables involved, to find
\begin{eqnarray}
{{u_1 r}\over{3g}}\,{}^{\#}\underline{d}\rho _{\rm m} &=& -\rho _{\rm m}\,{{h_0{}^0}\over{2 u^0}}\overline{\eta}_0\,,\label{gradcond01}\\
\bigl( k_1 -2\bigr){{u_1 r}\over{3g}}\,{}^{\#}\underline{d}n &=& -n\,{{h_0{}^0}\over{2 u^0}}\overline{\eta}_0 -{1\over 2}\bigl( k_1 -3\bigr)n\,u_1\overline{\eta}_1\,,\nonumber\\
\label{gradcond02}\\
{{u_1 r}\over{3g}}\,{}^{\#}\underline{d}\bigl(\beta T\bigr) &=& -\gamma T\,{{h_0{}^0}\over{2 u^0}}\overline{\eta}_0\nonumber\\
&&-{{\bigl( k_1 -3\bigr)}\over{\bigl(1 -b n\bigr)}} k_{_B} n T\,u_1\overline{\eta}_1\,,\label{gradcond03}\\
{{u_1 r}\over{3g}}\,{}^{\#}\underline{d}S_\mu{}^\mu &=& - S_\mu{}^\mu \,\Bigl( {{u_1^2}\over{2 u^0}}\overline{\eta}_0 + u_1\overline{\eta}_1\Bigr)\,.\label{gradcond04}
\end{eqnarray}
In analogy to (\ref{gradform02}), Eq.(\ref{rho03explic}) with (\ref{identmderlong}) and (\ref{mattemtrace01}) takes the form
\begin{equation}
{{2 u_1 r}\over{3g}}\,{\it l}_u \Bigl( {{3m}\over{4\pi r^3}}\Bigr) = \bigl(\rho + p\bigr) h_0{}^0 + S_\mu{}^\mu u_1^2 -2\bigl( k_1 -3 \bigr) p u_1^2 \,,\label{rho03reduc}
\end{equation}
from which we obtain the following conditions on the Lie derivatives
\begin{eqnarray}
{{r}\over{3g}}\,{\it{l}}_u\rho _{\rm m} &=& \rho _{\rm m}\,{{h_0{}^0}\over{2 u_1}}\,,\label{Liedercond01}\\
\bigl( k_1 -2\bigr){{r}\over{3g}}\,{\it{l}}_u n &=& n\,{{h_0{}^0}\over{2 u_1}} -{1\over 2}\bigl( k_1 -3\bigr)n\,u_1\,,\label{Liedercond02}\\
{{r}\over{3g}}\,{\it{l}}_u\bigl(\beta T\bigr) &=& \gamma T\,{{h_0{}^0}\over{2 u_1}} -{{\bigl( k_1 -3\bigr)}\over{\bigl(1 -b n\bigr)}} k_{_B} n T\,u_1\,,\nonumber\\
\label{Liedercond03}\\
{{r}\over{3g}}\,{\it{l}}_u S_\mu{}^\mu &=& - {1\over 2}S_\mu{}^\mu u_1\,.\label{Liedercond04}
\end{eqnarray}
The similitude between equations (\ref{gradcond01}) and (\ref{gradcond02}) or between (\ref{Liedercond01}) and (\ref{Liedercond02}) suggests to fix $k_1 =3$. In this case, (\ref{gradcond02}) and (\ref{gradcond03}) simplify respectively to
\begin{eqnarray}
{{u_1 r}\over{3g}}\,{}^{\#}\underline{d}n &=& -n\,{{h_0{}^0}\over{2 u^0}}\overline{\eta}_0\,,\label{gradcond05}\\
{{u_1 r}\over{3g}}\,{}^{\#}\underline{d}\bigl(\beta T\bigr) &=& -\gamma T\,{{h_0{}^0}\over{2 u^0}}\overline{\eta}_0\,,\label{gradcond06}
\end{eqnarray}
and the function (\ref{beta01}) reduces to $\beta = c_{\rm n} k_{_B} n$, so that, taking into account the definition (\ref{gamma01}), from (\ref{gradcond05}) and (\ref{gradcond06}) follows
\begin{equation}
{{u_1 r}\over{3g}}\,c_{\rm n}{}^{\#}\underline{d}T = -{1\over{\bigl( 1-bn\bigr)}} T\,{{h_0{}^0}\over{2 u^0}}\overline{\eta}_0
\,.\label{gradcond07}
\end{equation}
Eqs.(\ref{gradcond05}) and (\ref{gradcond07}) substituted into (\ref{enflux13}) (with $k_1=3$) allow to simplify the latter. However, it is easier to get the same result by using the heat part of (\ref{nogradenflow01}), that is
\begin{eqnarray}
\mathfrak{q}_{_H} &=& - T\sigma\Bigl[\,u^0\overline{\eta}_0 + h_0{}^0 \Bigl( {1\over{2 u^0}}\overline{\eta}_0 -{1\over{2 u_1}}\overline{\eta}_1\Bigr)\Bigr]\nonumber\\
&& + S_\mu{}^\mu u_1^2\,\Bigl( {1\over{2 u^0}}\overline{\eta}_0 +{1\over{2 u_1}}\overline{\eta}_1\Bigr)\,,\label{nogradheatflow01}
\end{eqnarray}
which, with the help of (\ref{gradcond04}) and (\ref{gradcond07}) transforms into
\begin{eqnarray}
\mathfrak{q}_{_H} &=& {{u_1 r}\over{3g}}\Bigl[\bigl( 1-bn\bigr) c_{\rm n}\sigma{}^{\#}\underline{d}T - {}^{\#}\underline{d}S_\mu{}^\mu\Bigr] -{1\over 2}S_\mu{}^\mu\,u_1\overline{\eta}_1\nonumber\\
&& - T\sigma\Bigl(\,u^0\overline{\eta}_0 -{{h_0{}^0}\over{2 u_1}}\overline{\eta}_1\Bigr)\,.\label{gradheatflow02}
\end{eqnarray}
Furthermore, with $k_1 =3$, (\ref{Liedercond02}) and (\ref{Liedercond03}) reduce to
\begin{eqnarray}
{{r}\over{3g}}\,{\it{l}}_u n &=& n\,{{h_0{}^0}\over{2 u_1}}\,,\label{Liedercond05}\\
{{r}\over{3g}}\,{\it{l}}_u\bigl(\beta T\bigr) &=& \gamma T\,{{h_0{}^0}\over{2 u_1}}\,,\label{Liedercond06}
\end{eqnarray}
implying
\begin{equation}
{{r}\over{3g}}\,c_{\rm n}\,{\it{l}}_u T = {1\over{\bigl( 1-bn\bigr)}} T\,{{h_0{}^0}\over{2 u_1}}\,.\label{Liedercond07}
\end{equation}
Replacing (\ref{Liedercond04}) and (\ref{Liedercond07}) in (\ref{gradheatflow02}), a third version of the heat flux is obtained
\begin{eqnarray}
\mathfrak{q}_{_H} &=& {{r}\over{3g}}\bigl( 1-bn\bigr) c_{\rm n}\sigma \Bigl( u_1{}^{\#}\underline{d}T + {\it{l}}_u T \overline{\eta}_1\Bigr)\nonumber\\
&& - {{r}\over{3g}}\Bigl( u_1{}^{\#}\underline{d}S_\mu{}^\mu -{\it{l}}_u S_\mu{}^\mu\overline{\eta}_1\Bigr) - T\sigma\,u^0\overline{\eta}_0\,.\nonumber\\
\label{gradheatflow03}
\end{eqnarray}
A similar treatment allows to get, from (\ref{nogradenflow01}) with (\ref{rhopluspdecomp01}), (\ref{gradcond01}), (\ref{Liedercond01}), (\ref{gradcond05}) and (\ref{Liedercond05}), three different forms for each of the remaining pieces of (\ref{energfluxdecomp02}), namely
\begin{eqnarray}
-\mathbb{J}_{\rm m} &=& -\rho _{\rm m}\Bigl[\,u^0\overline{\eta}_0 + h_0{}^0\Bigl( {1\over{2 u^0}}\overline{\eta}_0 -{1\over{2 u_1}}\overline{\eta}_1\Bigr)\Bigr]\,,\label{nogradmassflow01}\\
&=& {{u_1 r}\over{3g}}\,{}^{\#}\underline{d}\rho _{\rm m} -\rho _{\rm m}\Bigl(\,u^0\overline{\eta}_0 -{{h_0{}^0}\over{2 u_1}}\overline{\eta}_1\Bigr)\,,\label{gradmassflow02}\\
&=& {{r}\over{3g}}\Bigl( u_1{}^{\#}\underline{d}\rho _{\rm m} + {\it{l}}_u\rho _{\rm m}\,\overline{\eta}_1\Bigr) -\rho _{\rm m} u^0\overline{\eta}_0\,,\label{gradmassflow03}
\end{eqnarray}
and
\begin{eqnarray}
-\mathcal{J} &=& -n\Bigl[\,u^0\overline{\eta}_0 + h_0{}^0 \Bigl( {1\over{2 u^0}}\overline{\eta}_0 -{1\over{2 u_1}}\overline{\eta}_1\Bigr)\Bigr]\,,\label{nogradpartflow01}\\
&=& {{u_1 r}\over{3g}}\,{}^{\#}\underline{d}n - n\Bigl(\,u^0\overline{\eta}_0 -{{h_0{}^0}\over{2 u_1}}\overline{\eta}_1\Bigr)\,,\label{gradpartflow02}\\
&=& {{r}\over{3g}}\Bigl( u_1{}^{\#}\underline{d}n + {\it{l}}_u n\,\overline{\eta}_1\Bigr) - n u^0\overline{\eta}_0\,.\label{gradpartflow03}
\end{eqnarray}
Compare (\ref{gradmassflow02}) with (\ref{enflux03}) and (\ref{gradpartflow02}) with (\ref{enflux14}). Having the three pieces of (\ref{energfluxdecomp02}) basically the same structure, let us briefly comment the form (\ref{gradpartflow03}) of the particle number flow as a representative of the remaining contributions. Its first term is the gradient of the particle number density, assimilable to the local concentration of the only particle species considered, and thus representing the diffusive flux according to Fick's first law. (Similarly, the temperature gradient in (\ref{gradheatflow03}) tells about heat conduction in view of the Fourier law.) The last term in (\ref{gradpartflow03}), being $- n u^0\overline{\eta}_0 = n\bigl( u_1\overline{\eta}_1 + u_2\overline{\eta}_2 + u_3\overline{\eta}_3\bigr) = n u^a\overline{\eta}_a$ (where $u^a$ are, if not the components of a three-velocity in the strict sense, at least the spatially directed components of the four-velocity), is immediately interpretable as advective flux due to the fluid's main motion. Diffusion occurs relatively to this main stream, and the sum of the advective and the diffusive flows regarded together constitutes the convective flux. In addition to it, our result (\ref{gradpartflow03}) predicts the existence of a further contribution to the total energy flux, proportional to the change in time of the concentration $n$ as measured by its Lie derivative. As pointed out above, a similar interpretation holds for the flux pieces (\ref{gradheatflow03}) and (\ref{gradmassflow03}).

\subsection{Van der Waals complete energy-momentum}

The constitutive elements of $\Sigma ^{\rm VdW}_\alpha$ in (\ref{starem01}), with the general structure (\ref{symmenmom02}), are thus the following ones. The energy density $\rho$ consists of (\ref{rhodecomp01}) with (\ref{VdWgasenergydens02}), the pressure $p$ is given by (\ref{VdWgaspress02}), and the energy flux components are for instance (\ref{flux08})-(\ref{flux11}), whose decomposition into the three pieces (\ref{energfluxdecomp02}), each one with three alternative formulations, we have just shown. It remains to calculate the viscosity stress tensor, whose shear part (\ref{tracelessS02simplif}) is partially determined by replacing in it the already known expressions for density and pressure. Only the bulk viscosity is left out of the scheme, requiring to be measured by independent phenomenological methods. However, some information about it can be found from (\ref{gradcond04}) and (\ref{Liedercond04}), yielding for the total exterior derivative of $S_\mu{}^\mu$ the equation
\begin{equation}
{{r}\over{3g}}\,dS_\mu{}^\mu + S_\mu{}^\mu\Bigl(\vartheta ^1 -{{u_1}\over{2 u^0}}\vartheta ^0\Bigr) =0\,.\label{bulkder01}
\end{equation}
Taking into account (\ref{cofr01}) and (\ref{cofr02}), Eq.(\ref{bulkder01}) transforms into
\begin{equation}
{{r}\over{3}}\,d\log{S_\mu{}^\mu} + dr -{{u_1}\over{2 u^0}}fg\,dt =0\,,\label{bulkder02}
\end{equation}
or in more compact form, into
\begin{equation}
d\log{\bigl( r^3 S_\mu{}^\mu\bigr)} -{{3 u_1}\over{2 u^0}}{{fg}\over{r}}\,dt =0\,.\label{bulkder03}
\end{equation}
Integrating, we get the formal result
\begin{equation}
S_\mu{}^\mu  = {{K_{_S}}\over{r^3}}\exp\Biggl({{3}\over{2}}\int {{u_1}\over{u^0}}{{fg}\over{r}}\,dt\Biggr)\,.\label{bulkder04}
\end{equation}
The quantities $\rho _{\rm m}$, $n$ and $T$ are subject to similar integrable conditions. Actually, from (\ref{gradcond01}) and (\ref{Liedercond01}) follows
\begin{equation}
{{u_1 r}\over{3g}}\,d\rho _{\rm m} = \rho _{\rm m}\,{{h_0{}^0}\over{2 u^0}}\,\vartheta ^0\,,\label{rhomder01}
\end{equation}
so that
\begin{equation}
d\log\rho _{\rm m} = {{3 h_0{}^0}\over{2 u_1 u^0}}\,{{fg}\over{r}}\,dt\,,\label{rhomder02}
\end{equation}
implying
\begin{equation}
\rho _{\rm m} = K_{\rho _{_{\rm m}}}\exp\Biggl({{3}\over{2}}\int {{h_0{}^0}\over{u_1 u^0}}{{fg}\over{r}}\,dt\Biggr)\,.\label{rhomder03}
\end{equation}
Analogously, from (\ref{gradcond05}) and (\ref{Liedercond05}) we get
\begin{equation}
n = K_n\exp\Biggl({{3}\over{2}}\int {{h_0{}^0}\over{u_1 u^0}}{{fg}\over{r}}\,dt\Biggr)\,,\label{rhomder03}
\end{equation}
and on the other hand (\ref{gradcond07}) and (\ref{Liedercond07}) yield
\begin{equation}
{{u_1 r}\over{3g}}\,c_{\rm n}dT = {1\over{\bigl( 1-bn\bigr)}} T\,{{h_0{}^0}\over{2 u^0}}\,\vartheta ^0\,,\label{tempder01}
\end{equation}
from which it follows
\begin{equation}
T = K_{_T}\exp\Biggl({{3}\over{2\,c_{\rm n}}}\int {{h_0{}^0}\over{\bigl( 1-bn\bigr) u_1 u^0}}{{fg}\over{r}}\,dt\Biggr)\,.\label{tempder02}
\end{equation}
These results are consistence conditions of the present approach, relating the auxiliary variables $\rho _{\rm m}$, $n$ and $T$ to the gravitational functions $m$ and $\chi$ contained in $f$ and $g$, see (\ref{gfunct01}) and (\ref{ffunct02}). However, their physical meaning remains unclear.

\subsection{Photon gas energy-momentum}

Regarding the energy-momentum $\Sigma ^{\rm ph}_\alpha$ of thermal radiation in (\ref{starem01}), being photons massless, (\ref{rhodecomp01}) reduces to $\rho _{\rm ph} = \mathsf{u}_{\rm ph}$ so that the energy density coincides with the internal energy density (\ref{photgasenergydens01}), while pressure is given by (\ref{photgaspress01}). Let us assume as a general fact the vanishing of the energy-momentum tensor trace for arbitrary massless fields, in particular also for the photon gas
\begin{equation}
\bigl(\Sigma _{\rm ph}\bigr)_\mu{}^\mu = 0\,,\label{zerophemtrace01}
\end{equation}
in analogy to (\ref{ememtrace01}). Then, from (\ref{pressandS02}) with (\ref{photgasenergydens01}) and (\ref{photgaspress01}) follows
\begin{equation}
\bigl( S_{\rm ph}\bigr)_\mu{}^\mu = 0\,,\label{zerophemtrace01}
\end{equation}
and from (\ref{mattemtrace01}), with $k$ denoted as $k_2$, we find
\begin{equation}
{{3 m_{\rm ph}}\over{4\pi r^3}} = k_2 p_{\rm ph}\,.\label{mattemtrace02}
\end{equation}
The possibility of a non-zero contribution to the total gravitational mass due to massless particles derives from equation (\ref{exlicitmass01bis}) establishing the role played by internal energy at this respect. As a particular case, thermal photons contribute to the effective gravitational {\it mass} (as localized energy) as
\begin{equation}
m _{\rm ph}:= \int _0^r \mathsf{u}_{\rm ph}\,4\pi r^2 dr\,.\label{photmass01}
\end{equation}
Let us point out the following consistence condition. Taking (\ref{photgaspress01}) into account, from (\ref{mattemtrace02}) follows
\begin{equation}
m_{\rm ph} = {{4\pi k_2 }\over{9}}\,r^3\mathsf{u}_{\rm ph}\,,\label{photmass02}
\end{equation}
with radial derivative
\begin{equation}
\partial _r m_{\rm ph} = {{4\pi k_2 }\over{9}}\,\bigl( 3\,r^2\mathsf{u}_{\rm ph} + r^3\partial _r\mathsf{u}_{\rm ph}\bigr)\,.\label{photmass03}
\end{equation}
Comparing (\ref{photmass03}) with the derivative of (\ref{photmass01}), that is
\begin{equation}
\partial _r m _{\rm ph}:= 4\pi r^2\,\mathsf{u}_{\rm ph}\,,\label{photmass04}
\end{equation}
we get
\begin{equation}
\partial _r\log\mathsf{u}_{\rm ph} = -3\Bigl( 1 -{{3}\over{k_2}}\Bigr)\partial _r\log r\,,\label{photmass05}
\end{equation}
yielding
\begin{equation}
\mathsf{u}_{\rm ph} = \Phi (t)\,r^{-3\bigl( 1 -{{3}\over{k_{_2}}}\bigr)}\,.\label{photmass06}
\end{equation}
Comparing with (\ref{photgasenergydens01}), Eq.(\ref{photmass06}) informs about the radial dependence of the temperature as a function of the constant $k_2$.

On the other hand, the vanishing of the mass density implies that $\mathbb{J}_{\rm m}^{\rm ph} =0$, and due to (\ref{photgaschempot01}), we also find that $\mu _{\rm ph}\mathcal{J}^{\rm ph} =0$. Thus, according to (\ref{energfluxdecomp02}), the total energy flow of photons reduces to the heat contribution, for which we next obtain its three alternative formulations. From (\ref{nogradenflow01}) with (\ref{photgasenergydens01}), (\ref{photgaspress01}) and (\ref{zerophemtrace01}) follows
\begin{equation}
\mathfrak{q}_{_H}^{\rm ph} = -{4\over 3}\,\alpha T^4\Bigl[\,u^0\overline{\eta}_0 + h_0{}^0 \Bigl( {1\over{2 u^0}}\overline{\eta}_0 -{1\over{2 u_1}}\overline{\eta}_1\Bigr)\Bigr]\,.\label{photflow01}
\end{equation}
On the other hand, (\ref{enflow08}) with (\ref{zerophemtrace01}) and (\ref{mattemtrace02}) yields
\begin{eqnarray}
\mathfrak{q}_{_H}^{\rm ph} &=& {u_1 r\over{3g}}\,{}^{\#}\underline{d}\bigl( k_2 p_{\rm ph}\bigr) +\bigl(\mathsf{u}_{\rm ph} +p_{\rm ph}\bigr)\Bigl[ -u^0\overline{\eta}_0 + {{h_0{}^0}\over{2u_1}}\overline{\eta}_1\Bigr]\nonumber\\
&&+\bigl( k_2 -3\bigr)p_{\rm ph}u_1\overline{\eta}_1\,.\label{photflux03}
\end{eqnarray}
Replacing the internal energy density (\ref{photgasenergydens01}) and the pressure (\ref{photgaspress01}) we get
\begin{eqnarray}
\mathfrak{q}_{_H}^{\rm ph} &=& {4 u_1 r\over{9g}}\,k_2\alpha\,T^3\,{}^{\#}\underline{d}T + {4\over 3}\,\alpha T^4\Bigl( -u^0\overline{\eta}_0 + {{h_0{}^0}\over{2u_1}}\overline{\eta}_1\Bigr)\nonumber\\
&&+\bigl( k_2 -3\bigr){1\over 3}\,\alpha T^4 u_1\overline{\eta}_1\,,\label{photflux04}
\end{eqnarray}
Finally, taking into account (\ref{rho03reduc}) (with $k_2$ instead of $k_1$) as much as (\ref{zerophemtrace01}) and (\ref{mattemtrace02}), from (\ref{photflux03}) we find
\begin{eqnarray}
\mathfrak{q}_{_H}^{\rm ph} &=& {r\over{3g}}\,\Bigl[ u_1{}^{\#}\underline{d}\bigl( k_2 p_{\rm ph}\bigr) + {\it{l}}_u\bigl( k_2 p_{\rm ph}\bigr)\,\overline{\eta}_1\Bigr]\nonumber\\
&& -\bigl(\mathsf{u}_{\rm ph} +p_{\rm ph}\bigr) u^0\overline{\eta}_0 +2\bigl( k_2 -3\bigr)p_{\rm ph}u_1\overline{\eta}_1\,,\nonumber\\
\label{photflux05}
\end{eqnarray}
giving rise with (\ref{photgasenergydens01}) and (\ref{photgaspress01}) to
\begin{eqnarray}
\mathfrak{q}_{_H}^{\rm ph} &=& {4 r\over{9g}}\,k_2\alpha\,T^3\,\Bigl( u_1{}^{\#}\underline{d}T + {\it{l}}_u T\,\overline{\eta}_1\Bigr)\nonumber\\
&& -{4\over 3}\,\alpha T^4\,u^0\overline{\eta}_0 +\bigl( k_2 -3\bigr){2\over 3}\,\alpha T^4 u_1\overline{\eta}_1\,.\label{photflux06}
\end{eqnarray}
On its part, the shear viscous tensor (\ref{tracelessS02simplif}) with (\ref{photgasenergydens01}), (\ref{photgaspress01}) and (\ref{zerophemtrace01}) takes the form
\begin{equation}
\bigl({\slash\!\!\!\!S}_{\rm ph}\bigr)_\alpha{}^\beta = - {4\over 3}\,\alpha T^4\,\Bigl\{ h_\alpha{}^0 h_0{}^\beta + {{h_0{}^0}\over{2 u_1 u^0}}\,\Bigl( h_\alpha{}^1 h_0{}^\beta - h_\alpha{}^0 h_1{}^\beta\Bigr)\Bigr\}\,,\label{tracelessS02phot}
\end{equation}
completing the set of constitutive elements of the energy-momentum of thermal radiation.

\subsection{Determining the total energy dissipation rate $\Theta _{\rm tot}$ due to irreversible processes}

With all the necessary information at hand about the conserved energy-momentum (\ref{starem01}) of a gaseous radiating sphere (up to the bulk viscosity of the Van der Waals fluid), it only remains to calculate the quantity $\Theta _{\rm tot}$ in order to make explicit the physical content of the thermodynamic equations (\ref{firstlawtherm01}) and (\ref{secondlawthermsimpl}). In the absence of torsion and spin currents, for a single species of particles (\ref{Thetatot01}) reduces to
\begin{eqnarray}
\Theta _{\rm tot}\overline{\eta} &=& -\underline{D}u^\alpha\wedge S_\alpha{}^\beta\,\overline{\eta}_\beta -\underline{d}\mu\wedge\mathcal{J}\nonumber\\
&& + q_\alpha\,{\cal \L\/}_u u^\alpha\,\overline{\eta} +\mu\,\Theta ^{_{\rm N}}\overline{\eta} +\Theta _{\rm m}\,\overline{\eta}\,.\label{rhsfirstlawtherm01}
\end{eqnarray}
The covariant acceleration ${\cal \L\/}_u u^\alpha$ vanishes in view of (\ref{accel01}). The term $\underline{d}\mu\wedge\mathcal{J}$ as much as those having their origin in Eqs.(\ref{massflow02}) and (\ref{partfluxdecomp03}) as
\begin{eqnarray}
{\it l}_u ( \rho _{\rm m}\overline{\eta}) +\underline{d}\,\mathbb{J}_{\rm m} &=& -\Theta _{\rm m}\overline{\eta}\,,\label{massflow03}\\
{\it l}_u ( n\overline{\eta}) +\underline{d}\,\mathcal{J} &=& -\Theta ^{_{\rm N}}\overline{\eta}\,,\label{partfluxdecomp04}
\end{eqnarray}
derive from the Van der Waals fluid exclusively, since only it is massive and, on the other hand, the contribution of the photon number current is avoided by the vanishing of the corresponding chemical potential, see (\ref{photgaschempot01}). Regarding the first term in the rhs of (\ref{rhsfirstlawtherm01}), using (\ref{tracelessS02simplif}) we get
\begin{eqnarray}
&& -\underline{D}u^\alpha\wedge S_\alpha{}^\beta\,\overline{\eta}_\beta =\nonumber\\
&& = S_\mu{}^\mu\Bigl\{ -\underline{D}u_1\wedge\overline{\eta}_1 + {{u_1}\over{2 u^0}}\Bigl(\underline{D}u^0\wedge\overline{\eta}_1 - \underline{D}u_1\wedge\overline{\eta}_0\Bigr)\Bigr\}\nonumber\\
&& + \bigl(\rho +p\bigr)\Bigl\{\underline{D}u^0\wedge\overline{\eta}_0 - {{h_0{}^0}\over{2u_1 u^0}}\Bigl(\underline{D}u^0\wedge\overline{\eta}_1 - \underline{D}u_1\wedge\overline{\eta}_0\Bigr)\Bigr\}\,,\nonumber\\
\label{encons02}
\end{eqnarray}
whose explicit form for Van der Waals and photon gases is found by replacing in it (\ref{rhoplusp01}) with (\ref{VdWgasenergydens02}), (\ref{VdWgaspress02}), (\ref{photgasenergydens01}) and (\ref{photgaspress01}). The whole contribution of thermal radiation to (\ref{rhsfirstlawtherm01}) originates in (\ref{encons02}). Recalling (\ref{zerophemtrace01}), we calculate it to be
\begin{eqnarray}
\Theta ^{\rm ph}_{\rm tot}\overline{\eta} &=& {4\over 3}\,\alpha T^4\Bigl\{\underline{D}u^0\wedge\overline{\eta}_0\nonumber\\
&&\hskip1.5cm - {{h_0{}^0}\over{2u_1 u^0}}\Bigl(\underline{D}u^0\wedge\overline{\eta}_1 - \underline{D}u_1\wedge\overline{\eta}_0\Bigr)\Bigr\}\,.\nonumber\\
\label{encons02}
\end{eqnarray}
This result shows that thermal radiation as described by (\ref{photgasenergydens01})-(\ref{photnumb01}) may be involved in irreversible processes, contrarily to the usual assumption $\Theta ^{\rm ph}_{\rm tot} = 0$ corresponding to a photon gas in equilibrium \cite{Prigogine}. The fulfillment of condition $\Theta _{\rm tot}\geq 0$ by (\ref{rhsfirstlawtherm01}) remains an open question.

\section{Final remarks}

All elements of (\ref{starem01}) have been determined with the help of the auxiliary quantities (\ref{VdWgasenergydens02})-(\ref{VdWgaschempot02})  (\ref{photgasenergydens01})-(\ref{photnumb01}) ensuring the compatibility between the state equations and the Gibbs equation (\ref{uLiederiv02}). We also have identified the separate pieces of the energy flux (\ref{energfluxdecomp02}) and the total energy (\ref{rhsfirstlawtherm01}) dissipated in irreversible processes. Then, the conservation of energy-momentum derived in the framework of PGT as the Noether identity (\ref{sigmamattconserv}) (simplified to (\ref{sigmamattconservred})) ensures the fulfillment of both, the laws of thermodynamics --since the energy conservation law (\ref{firstlawtherm01}) together with the Gibbs equation (\ref{uLiederiv02}) gives rise to the second law (\ref{secondlawthermsimpl})-- and the dynamical Navier-Stokes equations (\ref{redmattender05}) for momentum conservation. The latter ones involve $q_\alpha$ and $S_\alpha{}^\beta\overline{\eta}_\beta$. The quantity $q_\alpha$ regarded as momentum density can be used in any of its forms found above, being (\ref{flux08})-(\ref{flux11}) the simplest one, without gradients nor Lie derivatives, while the viscosity contribution is found from (\ref{tracelessS02simplif}) to be
\begin{eqnarray}
S_\alpha{}^\beta \overline{\eta}_\beta &=& S_\mu{}^\mu\,\Bigl\{ h_\alpha{}^1\,\overline{\eta}_1
+ {{u_1}\over{2 u^0}}\Bigl( h_\alpha{}^1\,\overline{\eta}_0 -h_\alpha{}^0\,\overline{\eta}_1\Bigr)\Bigr\}\nonumber\\
&& -\bigl(\rho +p\bigr)\Bigl\{ h_\alpha{}^0\,\overline{\eta}_0 +{{h_0{}^0}\over{2 u_1 u^0}}\,\Bigl( h_\alpha{}^1\,\overline{\eta}_0 -h_\alpha{}^0\,\overline{\eta}_1\Bigr)\Bigr\}\,.\nonumber\\
\label{viscous3form}
\end{eqnarray}
The dynamical equations complete the physical description of a spherical gaseous non-rotating body (constituting a simple model of a star) composed of a Van der Waals fluid emitting radiation due to thermal photons and generating a gravitational field proportional to the total gravitational mass (\ref{exlicitmass01bis}) including contributions of the internal energy and thus depending on temperature. Our results are immediately generalizable to other gravitational sources.

\section{Conclusions}

Conservation of energy-momentum as the gauge current of local spacetime translations falls within the scope of gauge theories of gravity, in particular PGT. The link to thermodynamics is shown up by a few supplementary assumptions concerning the splitting into separate energy and momentum conservation equations and the introduction of auxiliary variables such as the internal energy as a functional of entropy and other quantities. Gravitation, dynamics and thermodynamics reveal themselves as different aspects of a comprehensive view on macroscopic matter, as they are subject to a common unifying symmetry principle interweaving the parts of the whole. In the words of Heraclitus of Ephesus (540 to 480 B.C.): {\it It is wise, listening not to me but to the logos, to agree that all things are one} (D.K. 50). According to him, the world shall be conceived holistically, as a formally structured unity spread out in multiplicity, rather than as an articulated plurality built out of mutually connected but autonomously existing elements: {\it From all things one and from one thing all} (D.K. 10).  Disjointedness of any part or feature of the world is disregarded. Everything depends on everything inside the unity of the whole: {\it The wise is one, knowing the plan how all things are steered through all} (D.K. 41) \cite{Kahn}. My proposal is an attempt to actualize Heraclitus' old sentences on reality and its knowledge.




\begin{acknowledgments}
I wish to thank Profs. Friedrich Wilhelm Hehl and Alfredo Tiemblo for their lessons and constant encouragement along time.
\end{acknowledgments}

\appendix
\section{Eta basis and its foliation}

Condition (\ref{dualitycond}) in 4-dimensional spacetime makes it possible to use tetrads $\{\vartheta ^\beta\}$ as the dual coframe of the local Lorentz frame $\{e_\alpha\}$. Accordingly, arbitrary $p$-forms $\alpha$ can be written as
\begin{equation}
\alpha ={1\over{p\,!}}\,\vartheta ^{\alpha _1}\wedge
...\wedge\vartheta ^{\alpha _p}\,(e_{\alpha _p}\rfloor ...
e_{\alpha _1}\rfloor\alpha\,)\,,\label{pform}
\end{equation}
showing the universal non-minimal coupling of tetrads (that is, of nonlinear translational conections) to any physical quantity represented as a differential form. The Hodge dual of (\ref{pform}), namely
\begin{equation}
\,{}^*\alpha ={1\over{p\,!}}\,\eta ^{\alpha _1 ... \alpha _p}\,(e_{\alpha _p}\rfloor ... e_{\alpha _1}\rfloor\alpha\,)\,,\label{dualform}
\end{equation}
is expressed in terms of the eta basis defined as the Hodge dual of products of tetrads as follows
\begin{eqnarray}
\eta &:=&\,^*1 ={1\over{4!}}\,\eta _{\alpha\beta\gamma\delta}\,\vartheta ^\alpha\wedge\vartheta ^\beta\wedge\vartheta ^\gamma\wedge\vartheta ^\delta\,,\label{eta4form}\\
\eta ^\alpha &:=&\,^*\vartheta ^\alpha ={1\over{3!}}\,\eta ^\alpha{}_{\beta\gamma\delta} \,\vartheta ^\beta\wedge\vartheta ^\gamma\wedge\vartheta ^ \delta\,,\label{antisym3form}\\
\eta ^{\alpha\beta}&:=&\,^*(\vartheta ^\alpha\wedge\vartheta ^\beta\,)={1\over{2!}}\,\eta ^{\alpha\beta}{}_{\gamma\delta}\,\vartheta ^\gamma\wedge\vartheta ^\delta\,,\label{antisym2form}\\
\eta ^{\alpha\beta\gamma}&:=&\,^*(\vartheta ^\alpha\wedge\vartheta ^\beta\wedge\vartheta ^\gamma\,)=\,\eta ^{\alpha\beta\gamma}{}_\delta\,\vartheta ^\delta\,,\label{antisym1form}
\end{eqnarray}
where
\begin{equation}
\eta ^{\alpha\beta\gamma\delta}:=\,^*(\vartheta ^\alpha\wedge\vartheta ^\beta\wedge\vartheta ^\gamma\wedge\vartheta ^ \delta\,)\label{levicivita}
\end{equation}
is the Levi-Civita antisymmetric object, and (\ref{eta4form}) the four-dimensional volume element. Foliation of p-forms (\ref{pform}) is given by (\ref{foliat1}). Provided the local constant Minkowski metric with signature $o_{\alpha\beta}= diag(-+++)$ is chosen to raise and lower indices, then, on the one hand
\begin{equation}
{}^*{}^*\alpha = \bigl( -1\bigr)^p\,\alpha\,,\label{explgren}
\end{equation}
and on the other hand, foliation of the Hodge dual forms (\ref{dualform}) reads
\begin{equation}
{}^*\alpha =\,(-1)^p\, d\tau\wedge {}^{\#}\underline{\alpha} - {}^{\#}\alpha _{\bot}\,,\label{foliat2}
\end{equation}
being $^\#$ the Hodge dual operator defined on the three-dimensional spatial sheets, affecting the pieces of (\ref{foliat1}). Regarding the eta basis, its foliated form can be found by using (\ref{tetradfoliat}) to calculate the products
\begin{equation}
\vartheta ^\alpha\wedge\vartheta ^\beta = d\tau\,\Bigl( u^\alpha\,\underline{\vartheta}^\beta - u^\beta\,\underline{\vartheta}^\alpha \Bigr) + \underline{\vartheta}^\alpha\wedge\underline{\vartheta}^\beta\,,\label{teth02}
\end{equation}
etc. in the rhs of (\ref{eta4form})--(\ref{levicivita}). In this way one gets
\begin{eqnarray}
\eta &=& d\tau\wedge\overline{\eta}\,,\label{eta04}\\
\eta ^\alpha &=& -d\tau\wedge\overline{\eta}^\alpha - u^\alpha\,\overline{\eta}\,,\label{eta03}\\
\eta ^{\alpha\beta} &=& d\tau\wedge\overline{\eta}^{\alpha\beta} -\Bigl( u^\alpha\,\overline{\eta}^\beta - u^\beta\,\overline{\eta}^\alpha \Bigr)\,,\label{eta02}\\
\eta ^{\alpha\beta\gamma} &=& -d\tau\,\epsilon ^{\alpha\beta\gamma} -\Bigl( u^\alpha\,\overline{\eta}^{\beta\gamma} + u^\gamma\,\overline{\eta}^{\alpha\beta}
+ u^\beta\,\overline{\eta}^{\gamma\alpha}\Bigr)\,,\nonumber\\
\label{eta01}\\
\eta ^{\alpha\beta\gamma\delta}&=& -\Bigl( u^\alpha\,\epsilon ^{\beta\gamma\delta}-u^\delta\,\epsilon ^{\alpha\beta\gamma}+ u^\gamma\,\epsilon ^{\delta\alpha\beta}- u^\beta\,\epsilon ^{\gamma\delta\alpha}\Bigr)\,,\nonumber\\
\label{eta00}
\end{eqnarray}
with definitions
\begin{eqnarray}
\overline{\eta}&:=& \Bigl( u\rfloor \eta \Bigr) ={1\over{3!}}\,\epsilon _{\alpha\beta\gamma}\, \underline{\vartheta}^\alpha\wedge\underline{\vartheta}^\beta\wedge\underline{\vartheta}^\gamma ={}^{\#}1 \,,\label{3deta06}\\
\overline{\eta}^\alpha &:=&-\Bigl( u\rfloor \eta ^{\alpha}\Bigr) ={1\over{2!}}\,\epsilon ^\alpha{}_{\beta\gamma}\,\underline{\vartheta}^\beta\wedge\underline{\vartheta}^\gamma ={}^{\#}\underline{\vartheta}^\alpha \,,\label{3deta07}\\
\overline{\eta}^{\alpha\beta}&:=&\Bigl( u\rfloor \eta ^{\alpha\beta}\Bigr) =\,\epsilon ^{\alpha\beta}{}_{\gamma}\,\underline{\vartheta}^\gamma ={}^{\#}(\underline{\vartheta}^\alpha\wedge\underline{\vartheta}^\beta\,)\,,\label{3deta08}\\
\epsilon ^{\alpha\beta\gamma}&:=&-\Bigl( u\rfloor \eta ^{\alpha\beta\gamma}\Bigr) =\,u_\mu\,\eta ^{\mu\alpha\beta\gamma}={}^{\#}(\underline{\vartheta}^\alpha\wedge\underline{\vartheta}^\beta\wedge\underline{\vartheta}^\gamma\,)\,.\nonumber\\
\label{3deta09}
\end{eqnarray}
The spatial 3-form (\ref{3deta06}) is the three-dimensional volume element. From (\ref{eta03}) follows that $\overline{\eta} = u^\alpha\,\eta _\alpha$. The contractions between tetrads and eta basis in four dimensions (see for instance \cite{Hehl:1995ue}), when foliated reduce to
\begin{eqnarray}
\underline{\vartheta}^\mu\wedge\overline{\eta}_\alpha &=& h^\mu{}_\alpha\,\overline{\eta}\,,\label{rel04bis}\\
\underline{\vartheta}^\mu\wedge\overline{\eta}_{\alpha\beta} &=& -h^\mu{}_\alpha\,\overline{\eta}_\beta +h^\mu{}_\beta\,\overline{\eta}_\alpha\,,\label{rel03bis}\\
\underline{\vartheta}^\mu\,\epsilon _{\alpha\beta\gamma} &=& h^\mu{}_\alpha\,\overline{\eta}_{\beta\gamma} +h^\mu{}_\gamma\,\overline{\eta}_{\alpha\beta} +h^\mu{}_\beta\,\overline{\eta}_{\gamma\alpha}\,,\label{rel02bis}\\
0 &=& -h^\mu{}_\alpha\,\epsilon _{\beta\gamma\delta} + h^\mu{}_\delta\,\epsilon _{\alpha\beta\gamma} -h^\mu{}_\gamma\,\epsilon _{\delta\alpha\beta} + h^\mu{}_\beta\,\epsilon _{\gamma\delta\alpha}\,.\nonumber\\
\label{rel01bis}
\end{eqnarray}
Taking (\ref{dualitycondbisbis}) into account, we also find
\begin{eqnarray}
e_\alpha\rfloor\overline{\eta}&=&\overline{\eta}_\alpha\,,\label{contract02}\\
e_\alpha\rfloor\overline{\eta}_{\beta}&=&\overline{\eta}_{\beta\alpha}\,,\label{contract03}\\
e_\alpha\rfloor\overline{\eta}_{\beta\gamma}&=&\epsilon _{\beta\gamma\alpha}\,.\label{contract04}
\end{eqnarray}
In view of definition (\ref{3deta09}), the contraction of all objects (\ref{3deta07})-(\ref{3deta09}) with $u_\alpha$ vanishes. From (\ref{3deta07}) then follows that $0=u_\alpha\,\overline{\eta}^\alpha  ={}^{\#}(u_\alpha\,\underline{\vartheta}^\alpha )\,$, thus implying $u_\alpha\,\underline{\vartheta}^\alpha =0\,$. Finally, comparing the variations of (\ref{pform}) with those of (\ref{dualform}), one gets the relation
\begin{equation}
\delta \,{}^*\alpha =\,{}^*\delta\alpha -{}^*\left(\delta\vartheta ^\alpha\wedge e_\alpha\rfloor\alpha\,\right) +\delta\vartheta ^\alpha\wedge\left( e_\alpha\rfloor {}^*\alpha\,\right)\,,\label{dualvar}
\end{equation}
showing that the variation of forms affected by the Hodge star operator involve variation of the tetrads. This fact reveals to be relevant for calculating the energy-momentum (\ref{definition03b}) from Lagrangian densities such as (\ref{emLagrang1}), (\ref{DiracLagrang1}) etc. depending on physical fields coupled non minimally to (nonlinear) translative connections.

\section{Other formulas used in the main text}

Definition (\ref{Liederdef}) of Lie derivatives can be generalized by replacing ordinary exterior differentials by covariant ones \cite{Hehl:1995ue}. For instance, the covariant Lie derivative of tetrads is defined as
\begin{eqnarray}
{\cal \L\/}_u\vartheta ^\alpha &:=& D\left( u\rfloor\vartheta ^\alpha\right) + u\rfloor D\vartheta ^\alpha\nonumber\\
&=& D u^\alpha + T_{\bot}^\alpha\,,\label{thetaLiederiv01}
\end{eqnarray}
where $T_{\bot}^\alpha$ is the longitudinal part of the torsion (\ref{torsiondef}) when foliated as
\begin{equation}
T^\alpha = d\tau\wedge T_{\bot}^\alpha + \underline{T}^\alpha\,.\label{torsdec01}
\end{equation}
For covariant derivatives of tensor-valued p-forms, say $D\alpha ^\mu$, Eq.(\ref{derivfoliat}) generalizes to
\begin{equation}
D\alpha ^\mu = d\tau\wedge\bigl( {\cal \L\/}_u \underline{\alpha}^\mu -\underline{D}\alpha_\bot^\mu\bigr) +\underline{D}\,\underline{\alpha}^\mu\,.\label{covderfol01}
\end{equation}
Foliation of (\ref{thetaLiederiv01}) yields
\begin{equation}
{\cal \L\/}_u\vartheta ^\alpha =d\tau\,{\cal \L\/}_u u^\alpha + {\cal \L\/}_u\underline{\vartheta}^\alpha\,,\label{thetaLiederiv02}
\end{equation}
where
\begin{eqnarray}
{\cal \L\/}_u u^\alpha &:=& u\rfloor D u^\alpha\,,\label{thetaLiederiv03}\\
{\cal \L\/}_u\underline{\vartheta}^\alpha &:=& \underline{D} u^\alpha + T_{\bot}^\alpha\,,\label{thetaLiederiv04}
\end{eqnarray}
being (\ref{thetaLiederiv03}) the covariant acceleration. Let us find an expression of the Lie derivative of the volume element (\ref{3deta06}) useful for calculations. The covariant derivative of (\ref{antisym3form}) is
\begin{equation}
D\eta _\alpha = \eta _{\alpha\beta}\wedge T^\beta\,.\label{torsder01}
\end{equation}
The lhs of (\ref{torsder01}) found from (\ref{eta03}) with (\ref{covderfol01}) reads
\begin{equation}
D\eta _\alpha = -d\tau\wedge\Bigl[ {\cal \L\/}_u \bigl( u_\alpha \overline{\eta}\bigr) -\underline{D}\overline{\eta}_\alpha \Bigl]\,,\label{etader01}
\end{equation}
while the rhs can be expanded with the help of (\ref{eta02}) and (\ref{torsdec01}), so that (\ref{etader01}) transforms into
\begin{equation}
{\cal \L\/}_u \bigl( u_\alpha \overline{\eta}\bigr) -\underline{D}\overline{\eta}_\alpha = u_\alpha T_{\bot}^\beta\wedge\overline{\eta}_\beta - u_\beta T_{\bot}^\beta\wedge\overline{\eta}_\alpha -\overline{\eta}_{\alpha\beta}\wedge\underline{T}^\beta\,.\label{etader02}
\end{equation}
By contracting (\ref{etader02}) with $u^\alpha$, and taking (\ref{thetaLiederiv04}) into account, we get the Lie derivative of the spatial volume
\begin{equation}
{\it l}_u \overline{\eta} = {\cal \L\/}_u \underline{\vartheta}^\alpha\wedge\overline{\eta}_\alpha\,.\label{volLiederiv02}
\end{equation}
A similar treatment of $D\eta _{\alpha\beta} = \eta _{\alpha\beta\gamma}\wedge T^\gamma$, with the help of (\ref{eta02}), (\ref{eta03}) and (\ref{torsdec01}), yields
\begin{eqnarray}
h_\alpha{}^\beta {\cal \L\/}_u\overline{\eta}_\beta &=& -\overline{\eta}_{\alpha\beta}\wedge {\cal \L\/}_u \underline{\vartheta}^\beta\,,\label{surfder01}\\
h_\alpha{}^\beta \underline{D}\,\overline{\eta}_\beta &=& \overline{\eta}_{\alpha\beta}\wedge\underline{T}^\beta\,.\label{surfder02}
\end{eqnarray}
Reformulating (\ref{surfder02}) as
\begin{equation}
\underline{D}\,\overline{\eta}_\alpha = u_\alpha {\it l}_u \overline{\eta} - u_\alpha T_{\bot}^\beta\wedge\overline{\eta}_\beta +\overline{\eta}_{\alpha\beta}\wedge\underline{T}^\beta\,,\label{surfder02bis}
\end{equation}
and substituting (\ref{surfder02bis}) into (\ref{etader02}) regarding (\ref{contract02}), it follows
\begin{equation}
{\cal \L\/}_u u_\alpha + \bigl( e_\alpha\rfloor T_{\bot}^\mu\bigr) u_\mu =0\,,\label{accel01}
\end{equation}
consistently with the fact that, being $u_\beta\underline{\vartheta}^\beta =0$ as read out from (\ref{transtetrad}), necessarily
\begin{equation}
0= e_\alpha\rfloor {\it l}_u \bigl( u_\beta\underline{\vartheta}^\beta\bigr) = {\cal \L\/}_u u_\alpha + \bigl( e_\alpha\rfloor T_{\bot}^\beta\bigr) u_\beta \,.\label{accel02}
\end{equation}
That is, covariant acceleration with respect to the local (comoving) Lorentz frame considered here only differs from zero if torsion exists.

\end{document}